\pdfoutput=1 
\documentclass[acmsmall]{acmart}

\AtBeginDocument{%
  \providecommand\BibTeX{{%
    \normalfont B\kern-0.5em{\scshape i\kern-0.25em b}\kern-0.8em\TeX}}}
\usepackage{enumitem}

\setcopyright{acmcopyright}
\copyrightyear{2018}
\acmYear{2018}
\acmDOI{10.1145/1122445.1122456}

\acmJournal{JACM}
\acmVolume{37}
\acmNumber{4}
\acmArticle{111}
\acmMonth{8}

\begin{document}
\setcopyright{acmlicensed}
\acmJournal{PACMHCI}
\acmYear{2021} \acmVolume{5} \acmNumber{CSCW1} \acmArticle{157} \acmMonth{4} \acmPrice{15.00}\acmDOI{10.1145/3449231}
\title{"Facebook Promotes \textit{More} Harassment": Social Media Ecosystem, Skill and Marginalized Hijra Identity in Bangladesh}

\author{Fayika Farhat Nova}
\affiliation{%
  \institution{Marquette University}
}\email{fayikafarhat.nova@marquette.edu}

\author{Michael Ann DeVito}
\affiliation{%
  \institution{Northwestern University}
  }
\email{devitom@u.northwestern.edu}

\author{Pratyasha Saha}

\affiliation{%
  \institution{University of Dhaka}
}\email{pratyasha.saha1195@gmail.com}

\author{Kazi Shohanur Rashid}

\affiliation{%
  \institution{University of Liberal Arts Bangladesh}
}\email{shohanur.rashid.bba@ulab.edu.bd}

\author{Shashwata Roy Turzo}

\affiliation{%
  \institution{East West University}
}\email{royturzo2712@gmail.com}

\author{Sadia Afrin}

\affiliation{%
  \institution{Daffodil International University}
}
\email{sadia10-1496@diu.edu.bd}

\author{Shion Guha}

\affiliation{%
  \institution{Marquette University}
}
\email{shion.guha@marquette.edu}

\renewcommand{\shortauthors}{Nova and DeVito, et al.}
\renewcommand{\shorttitle}{}
\begin{abstract}
Social interaction across multiple online platforms is a challenge for gender and sexual minorities (GSM) due to the stigmatization they face, which increases the complexity of their self-presentation decisions. These online interactions and identity disclosures can be more complicated for GSM in non-Western contexts due to consequentially different audiences and perceived affordances by the users, and limited baseline understanding of the conflation of these two with local norms and the opportunities they practically represent. Using focus group discussions and semi-structured interviews, we engaged with 61 \textit{Hijra} individuals from Bangladesh, a severely stigmatized GSM from south Asia, to understand their overall online participation and disclosure behaviors through the lens of personal social media ecosystems. We find that along with platform audiences, affordances, and norms, participant skill/knowledge and cultural influences also impact navigation through multiple platforms, resulting in differential benefits from privacy features. This impacts how Hijra perceive online spaces, and shape their self-presentation and disclosure behaviors over time.
\\
\textbf{Content Warning:} This paper discusses graphic contents (e.g. rape and sexual harassment) related to Hijra.
\end{abstract}



\maketitle

\section {Introduction}
\begin{quote}
They raped me\dots multiple times. When I tried to file complaint against the rapists in the police station, the police officers raped me again. They said they were doing me a favor, as I would never get a respected man because of my identity. I was stuck within those vicious circle of sexual harassment until finally, I decided to share these experiences on social media expecting to be heard and saved (P2, 20)
\end{quote}
This is the lived experience of P2, who is a Hijra from Bangladesh and a participant in this study.
Hijra, who are widely referred to as "third gender" \cite{khan2009living} individuals, are a group of people in south Asia who do not conform to binary notions of male or female gender but rather combine or move between them \cite{khan2009living}\footnote{As all participants in this study referred to themselves as Hijra, we will use the term "Hijra" to denote individual(s) from Hijra community, and "hijra" to refer their gender.}. Hijra are stigmatized and excluded from the society because of their perceived gender identities \cite{hossain2017paradox}, and many Hijra, like P2, experience extreme social exclusion, discrimination, harassment, and violence, with little or no access to physical, mental or social support. Instead, Hijra turn to social media for self expression, and social support.

Social media can play an amplified role for stigmatized populations, especially those with little access to physical assistance, including LGBTQ+ communities \cite{haimson2015disclosure, walker2020more, blackwell2016lgbt}. For such communities, social media acts as a primary space for identity exploration and development \cite{walker2020more, blackwell2016lgbt}, a primary source of social support and justice against harassment \cite{nova2019online, nova2019sharing, andalibi2017sensitive,andalibi2018social}, a resource for combating stigmatization around mental health \cite{betton2015role}, and both a guide and public platform for experiences such as gender transition \cite{haimson2015disclosure}. All of these happen across a multi-platform social media ecosystem, with individuals moving their content and attention between multiple platforms based on audiences, affordances, and their perceptions of the overall "spaces" available on a platform \cite{devito2018too}. Prior work \cite{devito2018too} has broken significant ground by exploring the importance of social media to members of gender and sexual minorities (GSM) in a Western, mostly US context. However, in order to improve and include social media platforms for everyone, we must better understand the need for and use of these platforms by stigmatized users in markedly different cultural contexts. Similarly, as we continue to improve the field's understanding and treatment of gender in platform design, it is essential to account for non-Western conceptions of gender and the needs and behavior of non-Western gender minority groups, such as Hijra, that may not necessarily align with the cultural or practical realities of Western GSM individuals. 

To address these concerns, we engaged with Hijra populations from Bangladesh and explored how GSM from non-Western contexts participate and self-present on different social media platforms, using DeVito et al.'s personal social media ecosystem framework for LGBTQ+ populations as a theoretical lens of inquiry \cite{devito2018too}. We find:
\begin{itemize}[noitemsep,topsep=0pt]
    \item Hijra primarily rely on social media platforms for three reasons: (a) communication with family, (b) Hijra community participation, and (c) sex work. Depending on each of these purposes, Hijra share content to targeted online audiences, as motivated by the platform's afforded levels of presentation flexibility and visibility control. 
    \item Technical knowledge and skill is a major factor in enabling Hijra to navigate social media platforms, with widespread lack of skill negatively impacting the way Hijra perceive platform affordances. Skill, when added to the input and influence of local authority figures (such as Hijra matriarchal leaders known as Gurumas), also motivates shifts in content across personal ecosystems and/or limitation of social media use.
    \item A reliance on Western cultural signifiers in designing platform features and navigation aids lessens the utility of social media for Hijra. Advanced and continually updating platform privacy features do not necessarily provide a sense of safety or practical benefit to Hijra if those features and their signifiers are not culturally familiar to them.
\end{itemize}
Whereas in prior work, GSM communities like LGBTQ+ users' online self-presentation and participation were analyzed through their audiences \cite{devito2018too}, platform affordances and the usability of the space, our paper adds to this conversation by finding that for GSM in non-Western contexts, this framework does not fully work. As the platforms' intended affordances to its users are not always aligned with Hijra's understanding of the platforms, for reasons like limited platform knowledge/skill or less culturally appropriated platform design, existing framework is unable to accurately explain how Hijra self-present themselves online with regards to their audiences and spaces. Previous literature has emphasized the importance of digital literacy within vulnerable communities in terms of their social media participation \cite{park2014understanding, park2013digital, sambasivan2018privacy}; however, such understanding is absent in the case of GSM from non-Western contexts. 

Therefore, our work makes several contributions to the CSCW community: 1) We extend and improve the current social media ecosystem framework \cite{devito2018too} by introducing and integrating technical knowledge and skill set in the framework based on the observations from Hijra community, 2) We contextualize the presence of community and cultural influence within Hijra groups, which helps us to better understand how GSM from non-Western contexts come to trust and prefer certain online platforms for their self-presentations and participation, 3) We advocate for design practices in HCI that integrate cultural context and marginalized views in the design phase to build more accurate, more inclusive social media environments for stigmatized GSM from non-Western context. Whereas existing work in ICTD discuss such inclusion and design practices from developing context \cite{medhi2009comparison, pal2013marginality, toyama2017design, kumar2019aspirations}, our study contributes to the conversation by including GSM populations like Hijra in HCI and CSCW.

\section{Background}

There is a growing scholarly recognition of the experience and diversity of sexual and gender orientations beyond binary gender and heterosexual identities \cite{bilodeau2005analysis}. Recent work in social computing has explored the benefits, pitfalls, and design opportunities around social media for GSM identities in a mostly US context \cite{ahmed2018trans, haimson2020designing, scheuerman2018safe, lerner2020privacy}. Similarly, researchers have begun to seriously grapple with the impacts of our concepts of gender on AI-based applications such as facial recognition in a mostly-western context \cite{keyes2018misgendering, scheuerman2019computers}. 
However, while these studies move us forward significantly, they ultimately categorize gender and sexuality through a strictly Western lens  \cite{bilodeau2005analysis, ibahmed_2017}. One sided perspectives can cause discrimination by disregarding the individual and cultural differences that these GSM populations experience throughout their lives. 
Due to the stigmatization and exclusion Hijra experience \cite{sandil2015negotiating}, their economic backwardness \cite{haque2019ulti}, their unique non-Western hijra identity, and their location in South Asia, 
Hijra are a crucial population to represent in order to broaden our understanding of GSM social media use in a non-western context.

As Hijra may be unfamiliar to many CSCW readers, we will first briefly illustrate the context in which Hijra use and rely on social media. We will then review specific concerns around self-presentation and digital literacy, and the overall personal social media ecosystem lens.

\subsection{Hijra: A History of Social Exclusion}
Hijra is an institutionalized third gender role that is neither male nor female, but contains elements of both \cite{nanda1986hijras}. Similar to Hijra, there are other gender minority identities that historically exist in many cultural contexts, known as \textit{bakla} in the Philippines, \textit{xaniths} in Oman, \textit{serrers} among the Pokot people of Kenya, and \textit{Hijra, jogappas, jogtas,} or \textit{shiv-shaktis} in South Asian countries such as, India, Bangladesh and Pakistan \cite{khan2009living, ibahmed_2017}. 
 Importantly, hijra is not simply a South Asian version of the Western concept of non-binary identity, a common misconception \cite{ibahmed_2017}. Hijra mostly live in segregated housing communes, where unwanted intersex or trans children are raised in a safer environment \cite{ibahmed_2017}. Hijra identity includes traditional procedures and distinct commitments unique to this form of gender minority identity, such as the time-honored ritual of leaving one's home — or being forced out — and undergoing induction into a clan of Hijra led by an elder matriarchal individual known as a "nayak" or \textit{Guruma}, at which point the new inductee is known as a "chela" or follower \cite{bearak_2016}. Gurumas and chelas have their own codes of conduct, and they often speak in Farsi-inflected variations of their local language known \textit{Ulti} to ensure secrecy \cite{bearak_2016}. 

Government estimates say there are around 10,000 Hijra in Bangladesh, although the "Badhan Hijra Songha", a transgender-Hijra rights group, states that the figure is actually around 100,000 \cite{wallen_2019}. Even though Hijra are legally recognized in Bangladesh, they are still socially excluded \cite{khan2009living}, with the word "Hijra" commonly used to mockingly refer to undesirable digression from normative masculinity  \cite{hossain2017paradox}. Despite formal recognition, this lingering lack of acceptance for gender identities beyond the binary results in limited employment opportunities for Hijra, many of whom turn to sex work or begging \cite{knight_2017}. Cisgender members of society often act to enforce the lower social status of Hijra, preventing them from accessing social institutions, resources, and services \cite{hossain2017paradox, khan2009living}. Hijra also face abnormally high rates of hate crimes involving rape, harassment and physical abuse \cite{goel2016Hijra}. Due to Hijra’s vulnerable social status and exclusions from the society, several non-governmental organizations (NGOs) try to support these communities for issues like sexual health, educations, job opportunities and so on \cite{aziz2019social}. NGOs such as Badhan Hijra Sangha are working hard to raise awareness of general and sexual well being among Hijra community; to create basic human, civic, and social rights for them and establish network with national, international GO/NGO /institutions \cite{aboutus_badhan}. They are also providing legal supports and skill development training for alternative livelihood options for Hijra to reduce economical discrimination \cite{aboutus_badhan}. Additionally, Sachetan Somajsheba Hijra Sangha, Bondhu Social welfare society, Hasab, Shomporker Naya Shetu, iccddr,b etc. are some of the notable NGOs in Bangladesh who have taken many initiatives for Hijra communities to establish group dynamics and advocacy with different level of stakeholders \cite{safa2016inclusion}.

Solidarity among Hijra is quite strong because they as a group are discriminated and excluded from the rest of society because of their specific cultural conventions and group norms that are not accepted by the normative populations \cite{khan2009living}. Even though dimensions of their social deprivation and harassment have never received attention in development sectors \cite{khan2009living}, it is time to change that notion in the CSCW research community, as we aim to reflect on our practices and design to be more inclusive and flexible towards all genders and sexualities.

\subsection{Online Self-presentation and Impression Management} 
Impression management involves the processes by which people control how they are perceived by others \cite{leary1990impression}. Social media plays an important role in how this impression or self-presentation is constructed for most users \cite{davis2017identity}. However, particularly for GSM, this management of impression becomes more critical, as it provides them opportunities to experiment with their self presentation and identity to the the world \cite{duguay2016he,kuper2014using}. Facebook and other social networking sites have been a major area of research, particularly to understand what practices and behaviors users adopt during their self-presentation online \cite{boyd2008youth, joinson2008looking, toma2015facebook}. Previous research has highlighted how social media can facilitate and assist LGBTQ+ users from Western context during self-disclosure through constructing, managing, and expressing identity projections \cite{bates2020let}. \cite{cooper2010facebook} investigates the use of Facebook by LGBTQ+ users to understand how identity construction, management and negotiation, and activism take place within these communities in a rural American social setting. 
The existing work on LGBTQ+ users identifies a variety of identity management strategies that these populations adopt on social media including monitoring their online self-expression, using privacy and security controls, strategically managing their audiences and so on in Western setting \cite{duguay2016he, cooper2010facebook, mcconnell2018everybody}. 

This process of constructing self-presentation decisions on distinct social media platforms becomes complicated due to LGBTQ+ users' complex and stigmatized identity \cite{devito2018people}. According to \cite{cooper2010facebook}, LGBTQ+ users struggle with their use of Facebook, as many of them are not open about their sexuality and gender identity and prefer to keep it as a secret on the online platforms by constructing a different, more acceptable identity than their actual self. 
While the vast majority of work on LGBTQ+ has been done from Western perspectives, there is some existing work that try to explore such GSM communities from a global South context. Studies like \cite{bacchetta1999hindu, gannon2007respect} have focused on queer and Hijra population from India to explore their realities in terms of social, economic, political, emotional, psychological, and legal issues. Indian LGBTQ+'s adoption of email lists, message boards, and weblogs to communicate with each other online have also been explored by researchers \cite{krishna2010sexuality, mitra2008queer, mitra2010resisting}. However, few studies seem to have focused upon the possible use of popular social networking platforms from Global South. 
Literature like \cite{dey2019queering} have added to that conversation by studying how Indian LGBTQ+ individuals create multiple identity on distinct social media, such as Facebook, to protect their gender minor identities from unwanted audiences. However, such exploration of GSM communities from developing context is understudied and can differ from how Hijra communities adapt to those online practices \cite{mcconnell2018everybody}. 
\begin{figure}
\centering
  \includegraphics[width=.6\columnwidth]{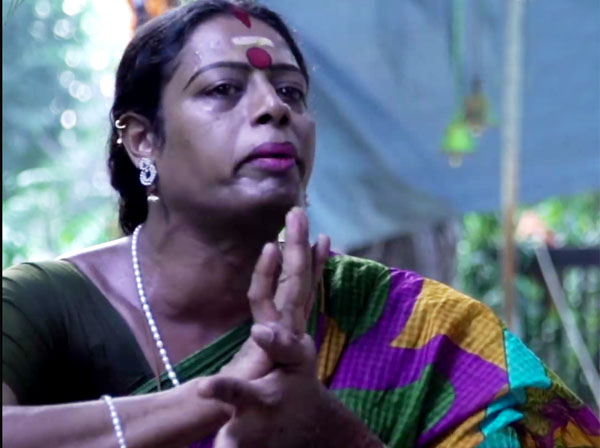}
  \caption{Tin Tali, traditional clap to demand attention or recognition by Hijra (\href{https://bit.ly/2U2qPqD}{Online photo})}~\label{fig:figure1}
  \vspace{-2em}
\end{figure}

Self-presentation and impression management is a crucial part of life for Hijra due to their stigmatized identity \cite{shawkat_2016}. In Bangladesh, where often talk on sex education is considered taboo \cite{deshwara_eagle_2017}, figuring out one's own gender and sexuality becomes a struggle, especially for the low-literate community such as Hijra \cite{ainul2017adolescents}. 
Some Hijra outwardly acknowledge that they are women through sexual relationships, gender roles, and clothing, but are aware that they are separate and distinct from cisgender women \cite{jayadeva2017understanding}. Different expectations from the gender roles, biological sex, and sexual orientation within Hijra can make it harder for them to cultivate their identity in the society \cite{Rep_hij_article}. As individuals shape their behavior for certain audiences in specific contexts \cite{goffman1978presentation}, many Hijra similarly continue to intentionally and surgically construct their self-presentation decisions for targeted audiences. Prior work on Hijra identity suggests that it is useful to view Hijra impression management through Goffman's dramaturgical model of self-presentation \cite{goffman1978presentation, bullingham2013presentation, dey2019queering}. Goffman's dramaturgical model posits that often individuals prefer to represent themselves differently to safe "backstage" audience regions and potentially-threatening "frontstage" audience regions \cite{goffman1978presentation, bullingham2013presentation}. It is purely contextual and up to the individual's need to decide which behavior or identity becomes the front stage. Many Hijra are forced to put on aggressive and hostile front on the streets as a defense mechanism to fight street harassment \cite{haque2019ulti, shawkat_2016, marichal2013political} even though in personal lives within their own community, they may have completely different personalities. Hijras can also be aggressive, especially when not handed money as they wend their way through traffic, begging as a primary means to earn money \cite{mccarthy2014journey} in a discriminatory society. Often these rude remarks are followed by \textit{tin tali} (a specific way of clapping by Hijra demanding for recognition of their existence \cite{bearak_2016} \textbf{Fig. 1}) and lifting up sarees showing their castrated genitals or breasts in public \cite{khan2009living}. This rude and sexualized presentation of Hijra in frontstage regions is a product of continuous repression and criticism they face in daily lives and can be very disconnected from their presentation in private, backstage regions. 
In direct contrast to presentation of Hijra in frontstage regions, their presentation in safe "backstage" regions is generally nicer, calmer and very friendly \cite{shawkat_2016}. Previous research has discussed this notion of front and backstage by exploring how individuals with stereotyped identities actively shape their online social identity through different self-presentation or disclosure strategies \cite{haimson2017social}. They often apply approaches of strategic outness, first assessing a specific social situation before determining whether to disclose their gender or sexuality online \cite{duguay2016he}. 
As disclosure or exploration of stigmatized identity on social media platforms can often lead to negative outcomes such as hate speech, name calling, and unsolicited sexual photos and messages \cite{blackwell2016lgbt, haimson2016transgender, haimson2016constructing, galupo2014disclosure}, it can effectively force users to restrict their online participation \cite{turan2013reasons, nova2019online} or employ heavily tailored privacy settings to manage visibility of their gender and sexuality related content within a single platform \cite{devito2018too}.

\subsection{Digital Literacy and Skill for Navigating Online Platforms}
One potentially major complicating factor around self-presentation behavior on social media platforms is lack of digital literacy or skill, as it enables users to choose effective technology practices to meet their goals \cite{park2013digital}. As the self-presentation goals of Hijra can be complex and may require use of advanced privacy features, lack of skill and knowledge of the platform features can be a major challenge for such stigmatized users \cite{hargittai2015mind}. According to previous research, privacy and security settings of social media platforms can often be difficult to navigate and imperfect in terms of user's requirement of information control \cite{mcconnell2018everybody}. For instance \cite{bartsch2016control} discusses how privacy literacy may change online behavior and perceived online safety within users; they define online privacy literacy as users' knowledge about technical aspects of online data protection, and ability to apply those strategies for own privacy regulation. Another previous study \cite{park2014understanding} has explored how different levels of privacy knowledge and skills among African American younger adults can effect their online practices. \cite{park2014understanding} has identified how populations with different level of skill and particular are consistently left out from benefits of technology because they cannot access the full potentials of the technology.


As an explanation for certain communities' limited skill to navigate online platforms, some studies suggested a combination of imperfect interface design and a lack of Internet literacy \cite{bartsch2016control, livingstone2007strategies}. \cite{park2013digital} also shows low online privacy literacy within the participants in terms of technical familiarity and policy knowledge. Technical skill and knowledge are often barriers for older populations who lack the understanding of rigorous privacy settings and features provided by the platform \cite{vitak2014you}. It has been also found that lower internet skill within women has consequently increased the gender gap on online spaces \cite{hargittai2015mind}.
Whereas these studies looked at the impact of technical skill and knowledge on populations from Western context, several studies found that it is more common within non-white users to fall behind in online privacy control behavior due to their limited skill set \cite{park2012affect} and knowledge \cite{park2013digital}. 
Even though previous research on digital literacy have not explored Hijra or other GSM's skill and knowledge of using online social media platforms from global South context, \cite{sambasivan2018privacy} talks about how less technical skill within women in south Asia impact and limit their technology use. Such exploration of different communities around the world directs our attention towards understanding the concept of digital literacy and knowledge within GSMs from non-Western context, such as Hijra. Due to Hijra's stigmatized identity, it is absolutely essential for them to utilize different platform affordances to ensure privacy and control over self-presentation settings, and thus, this study tries to fill that void of knowledge by exploring Hijra communities from Bangladesh.

\subsection{Social Media Ecosystem Framework}
In prior work researchers have conceptualized how GSM self-presentation, existing in across multi-platform ecosystem, has allowed differential presentation across different audiences and sites \cite{devito2018too}. 
Specifically, Devito et al. \cite{devito2018too} posited three specific elements of social media ecosystems that drive self-presentation behavior (and content) to appropriate outlets: audiences, affordances, and the conflation of the two with local norms, which they call "spaces".


\textit{Audience}. DeVito et al. \cite{devito2018too} found that different audience compositions per platform and, importantly, user perceptions of these audiences were a key motivator for personal social media ecosystem use and movement of content across said ecosystems \cite{devito2018too}. LGBTQ+ users generally conceptualized their audiences as either abstract (relatively unknown) or targeted (specific people who are the potential connections users may have and want to share contents with through their social media platforms). Users imagine their audience based on factors ranging from goals and individual psychological expectations from others, allowing them to act self-protectively despite rarely having access to precise audience composition information.

\textit{Affordances}. According to DeVito et al. \cite{devito2018too}, user perception of a platform and its appropriate place in one's personal social media ecosystem is heavily affected by the affordances, or possibilities for action, each platform offers to users . Stigmatized users explore and look beyond single platforms, considering the range of affordances available across their personal ecosystem when making self-presentation decisions.

\textit{Space}. DeVito et al. \cite{devito2018too} use the term "space" to describe the conflation of platform and audience by users making self-presentation decisions. Through the lens of this conflation, users form a concept of what "type" of platforms are available, what they are for, and who is welcome there. By examining the conflated spaces, we are able to look not just at the social context intended by platform designers, but rather what a platform represents to the users themselves in comparison to other platforms within one’s personal social media ecosystem.

DeVito et. al's framework of personal social media ecosystem \cite{devito2018too} provides a solid theoretical foundation for research into Hijra self-presentation and social media platform use. We draw from this theoretical perspective to frame our research questions and also as the qualitative lens for our analysis. However, although the existing framework provides valuable perspectives on stigmatized GSM populations in online spaces, it was developed entirely in a Western context and may need extension to apply to Hijra and other stigmatize populations from developing, non-Western contexts. To better account for Hijra and other non-Western gender minorities in the design of the social media platforms they rely on for crucial services and communication, it is imperative to extend this lens beyond a Western context. As such, using DeVito et al.'s personal social media ecosystems as a guide, we ask: 
\\
\\
\textbf{RQ1.} What are the audience related concerns that Hijra have in their social media ecosystem? 
\\
\textbf{RQ2.} How do they manage their audiences through the affordances of different social media platforms? 
\\
\textbf{RQ3.} How do audience and afforances influence Hijra to move around in different social media platforms?

\section{Methods}
To answer these questions, we conducted a six-month-long qualitative study of Hijra in Dhaka, Bangladesh from March-August 2019. The study was conducted in 6 neighborhoods of Dhaka \textbf{(see Fig. 2)}: Lalmatia, Kakrail, Mugdapara, Gulistan, Manda, and Kamalapur. During this period, we built trust and encouraged our participants to participate in the study by spending ample time in their slums \footnote{Here, slum refers to the colloquial reference in South Asian English to lower quality of informal housing without any derogatory connotation.} to meet with participants and built rapport. We employed multiple elicitation methods, including semi structured interviews, Focus Group Discussions (FGDs) \cite{nili2017framework} and unstructured online observations \cite{norskov2011observation} of public and private (with consent) content related to Hijra. To ensure a robust understanding of how Hijra interact with social media, we triangulated these multiple data sources in our analysis. All study procedures were approved by the Institutional Review Board at the lead author's academic institution, a private research university in the midwestern, United States.

\begin{figure}
\centering
  \includegraphics[width=.7\columnwidth]{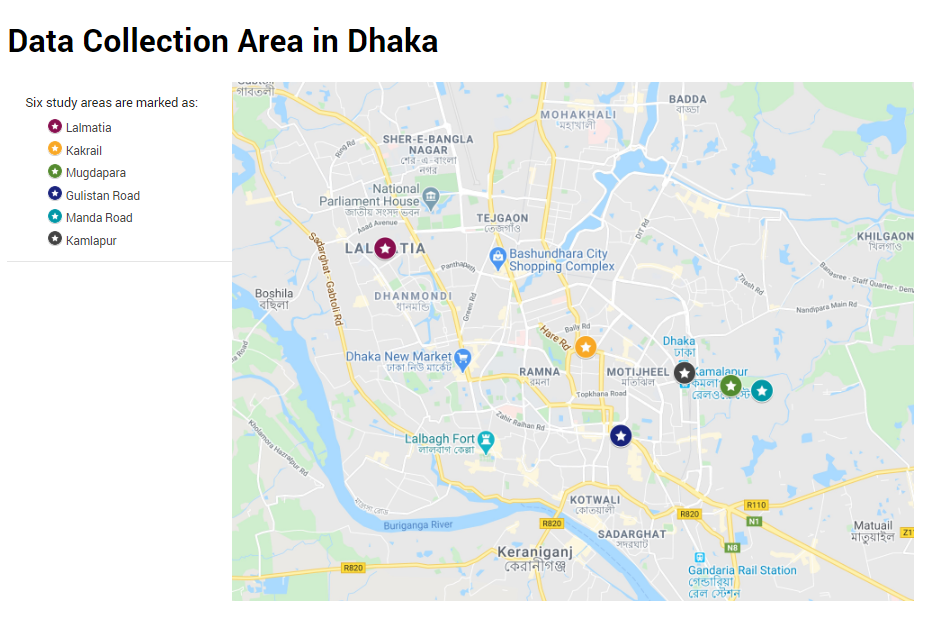}
  \caption{Field Areas for Data Collection in Dhaka}~\label{fig:figure2}
  \vspace{-2em}
\end{figure}
\subsection{Participants}
Participant recruitment was performed through a snowball-style iterative process of networking and trust building in Hijra communities. The third author, leading efforts on the ground, first conducted 3 preliminary semi-structured interviews with acquaintances who belong to Hijra community. From there, we were approved to conduct our first FGD consisting of 6 Hijra in a dorm at Lalmatia, where Hijras from different districts were gathered for a cultural event. Our second FGD also consisted of 6 participants, all Hijra sex workers who were visiting Dhaka for a training program arranged by a local NGO. We arranged a meeting with them with the help of \textit{Sachetan Somajsheba Hijra Sangha}, a non-profit organization working for the welfare of Hijra community. The remaining 3 FGDs were conducted with local Hijra in a place they considered both convenient and safe. 

Though the in-person activities in this study were conducted in Dhaka, we ensured representation from participants all over the country. We talked to Hijra communities from 8 divisions for the FGDs and individual interviews: Barisal(3 participants), Chittagong(4), Dhaka(37), Khulna(3), Mymensingh(3), Rajshahi(5), Rangpur(3), and Sylhet(3). In total we had 61 participants (45 from FGDs, 16 from one-one interviews). 
\textbf{Table 1} shows for additional demographic information.
Almost all participants were employed doing "Hijragiri" (traditionally, the ritual of \textit{badhai}, or blessings conferred on a newborn through dancing and singing) \cite{safa2016inclusion}, the collection of \textit{cholla} (tolls from jurisdictions), training to become skilled in the Ulti language \cite{haque2019ulti}, and sex work. Out of 61 participants, 19 were directly involved with sex work, and only 3 participants were involved with white-collar jobs. The socio-economic status of all the participants was relatively low, as Hijra community in Bangladesh conventionally consists of those from lower economic levels \cite{hijraarticle}. 

\subsection{Data Collection}
Data collection employed 5 Focus Group Discussions (FGDs), 16 one-on-one interviews, and online observations on participants' social media participation with their prior consent. Interviews and FGD data were collected in the form of field notes and audio recordings. We collected photos, screenshots, videos from our online observations. All the interviews and FGDs were conducted in Bengali, and each participant in the focus group and interview was compensated with BDT 400Tk, which is roughly around \$5 and more than the minimum daily wage (BDT 50TK) in Bangladesh \cite{minwage}. In all data collection activities, we focused on the social media practices of Hijra communities, with specific attention paid to audience management strategies and social media participation. 

\begin{table}
\begin{tabular}{lll}
\hline
\multicolumn{2}{c}{\textbf{Demographics}} & \multicolumn{1}{c}{\textbf{Percentage}} \\  
\hline
Sex By Birth & Male & 93\% \\ 
 & Intersex & 7\% \\ 
\hline
Sex Assigned at Birth & Male & 100\% \\ \hline
Gender & hijra & 100\% \\ \hline
Preferred Hijra Identity &  Hijra   & 47.6\% \\
     
& Third Gender  & 21.3\%\\

& Women & 31.1\%\\
\hline
Sexuality & Gay & 100\% \\ 
\hline
Age Range & 18-25 yrs & 65\% \\  
 & 26-33 yrs & 22\% \\  
 & yrs > 33  & 13\% \\ \hline
Highest Level of Education & No Education & 38\% \\ \ 
 & Primary Education & 24\% \\ 
 & Some High School & 17\% \\  
 & High School Diploma & 13\% \\ 
 & Bachelors Degree & 8\% \\ \hline
Location & From Dhaka & 87.43\% \\ \hline
\end{tabular}
\caption{Demographics of the Participants}
\label{tab:my-table}
\vspace{-2em}
\end{table}

Our initial data collection strategy centered around the 5 FGDs, each of which was 2.5 hours long. However, we noticed that often the most senior members of the group (such as Guruma) responded to our questions or took control of the room, as opposed to all the members being involved in the discussion. To balance this, we decided to conduct the one-on-one interviews with follower Hijra who live under Gurumas. The 16 one-to-one interviews were conducted with the help of a Guruma who is also a Hijra activist, and helped us reach more people to talk to individually. The time and date for these interviews were chosen according to the participants' preference. Interviews averaged 40 minutes each.

Our interview and FGD protocols consisted of 24 sets of questions, with multiple sub-questions under each of them. Answering every question was not compulsory, and the participants could skip questions if they wished. Later, the interviews were translated and transcribed by the 4th, 5th and 6th authors of this study, whose first language is Bengali. 

The online observations were collected through exploring different social media platforms that Hijra mentioned in FGDs and interviews, such as Facebook and Bigo Live. While looking at the social media information provided by our participants, we also used different keywords suggested by our participants, such as \textit{Hijra}, \textit{Third Gender} etc., to find online groups and communities that promote anti- or supportive posts related to Hijra. Keeping ethical implications in mind, we only collected information on groups and pages through key words that were already explicitly public for everyone on social media platforms \cite{fiesler2015ethics, bromseth2002public} and were not only accessible to only a certain community of users. As suggested by \cite{zimmer2010but}, it is unethical for researchers to use any personal information from social media if the data or information is restricted to a certain group of people or communities. Hence, we ensured the Facebook groups/pages we shared image from are public and not restricted to certain communities or populations. \cite{ravn2020publicly} also adds to this conversation of ethical implications by emphasizing that more than expected information should not be revealed through images’ background or through the combination of visual and textual elements. To ensure that, we intentionally refrained ourselves from publishing or revealing any post or personal profile that were shared in those groups even though they were public.

The FGDs and interviews were conducted in a secured setting, giving priority to participants' preference of time and place. All data were recorded only after the consent of participants, and names used in the writing of this paper are pseudonyms. Observational data from the participant social media account were only recorded with proper consent, and care was taken not to record participant names, profiles, or any kind of identifiable information. 

\subsubsection{Challenges in Data Collection}
Working with the Bangladeshi Hijra community was a complex process. Even though this study includes a high number of participants to understand Hijra's social media ecosystem, there were few challenges that our authors faced during data collection. The first challenge was to gain access to Hijra participants. Being largely disconnected from the civil society, Hijra are very protective towards their own community and hesitate to give outsiders access. It was even challenging to even commute to these communities for our author, as rickshaw pullers did not want to provide a trip to such locations, and uncooperative locals sometimes passed harassing comments to the author upon seeing her entering their places alone. Wishing to avoid the known problem of fake street Hijra \cite{khan_2016}, we initially intended to work through community organizations or NGOs to find participants. However, these organizations often did not ultimately cooperate due to privacy and funding issues. The second challenge was in actual data collection. Even though our author contacted Hijra through different organizations or NGOs, not everyone were interested in participating in the study. However, this number of non-participation was not high. Only 3 out of 64 Hijra we initially contacted declined to participate in the study because of their discomfort in sharing personal information on their gender, sexuality and social media participation. 
We respected their decision and only collected data from the rest 61 participants who were interested in the study. Additionally, 3 out of those 61 participants had some work conflicts where they received calls from their clients during FGDs and had to leave. In such cases, the data collection from those specific participants was stopped by the author with permission. We still included them in our dataset due to the depth and value of the data they provided and their sincere interest to be parts of the study. However, Hijra were not the primary complication we had to face during data collection; the middlemen who worked with us in the field were. As we had to go through NGOs to connect with the Hijra communities, sometimes they provided middle-men from their organization to coordinate and manage Hijra in the field (such as calling them one by one for the interview). Even though these individuals were sent as volunteers to assist our ethnographer in the field, without NGOs knowledge, these middlemen kept asking for a large amount of money as a bribe from our author with subtle threats of discontinuing the study otherwise. In a developing country like Bangladesh, corruption in the form of bribes is a huge concern \cite{desk_2019}. Despite such obstacles, we gave our best effort to continue the research and collected data from a large number of samples as majority of our participants were comfortable with the study and did not mind sharing their data. Finally, this project involved close observation of Hijra's day-to-day life while staying with them more than 6 hours a day. Due to the disturbing nature of some aspects of the day-to-day treatment of Hijra by society, team members found some parts of data collection personally upsetting, requiring later treatment by a therapist. We enumerate these issues in this paper as a guide for other researchers seeking to conduct similar research in challenging settings.

\subsection{Data Analysis}
After data collection, we anonymized the data before analysis. To anonymize, the first and third author renamed all the FGD participants as P1, P2... etc., and for interviews, as X1, X2... etc. As our data was mostly qualitative, we used a grounded, thematic approach \cite{zhang2009qualitative} on our collected data. For each source of data, the first author created codebooks following an open-coding approach to allow flexibility for new themes to emerge. We wanted to understand the rationale behind Hijra perceptions on gender identity, and online participation. The codebook was created through several iterative rounds of coding until we reached theoretical saturation \cite{glaser1968discovery}. The first author shared the codebook with other members of the team in each iteration and upon thorough discussion, the team reached agreement on the generated codes. The categories formed from the codes were later grouped into different themes which helped us construct our findings from this study. Although the FGDs provided us initial themes and observations on Hijra's group dynamics and overall experiences in their day-to-day and online lives, the interviews provided us more specific and nuanced information on Hijra. From the interviews, we drew deeper detail on their struggles, confusion, and frustration both in offline and online world, both confirming and expanding our initial FGD-based themes. The online observations helped us to provide visual references to the readers and connect the experiences of Hijra (that they shared through FGDs and interviews) with practical instances. 
 

\subsection{Author Positionality Statement}
Our study was a team effort, with varying specialties and expertise within the research team. Even though we set out with the intention to include a Hijra representative in our research team, there was a lack of interest from the Hijra community. Due to Hijra’s stigmatized social status, often these communities are supervised by different NGOs who regulate Hijra’s well-being and social participation. This results in a \textit{de facto} degree of social control between NGO and Hijra, which affects our access to these participants. Without NGO’s permission, no Hijra can directly participate in any study or research with any outside researchers due to possible conflicts or complexities that may arise, both political and social, hampering Hijra’s safety. Therefore, even though the activist Guruma assisted us in the paper, to ensure safety and anonymity, that Guruma did not want to be an author in our paper. 
In an effort to ensure a balanced expertise in the research team, our author group consists of cis and trans authors who work with GSM participants from both Global North and South. Three authors have previously worked with stigmatized and vulnerable populations, including Hijra, children and women, through prior engagement with different NGOs, one author is a transgender member-researcher for Western LGBTQ+ populations, and the remaining three team members have formal training on transcribing transcripts for academic research. Except the second author of this paper, who is American-born, the authors have native proficiency in speaking, writing, and understanding the Bengali language that Hijra used in this study; grew up in South Asia and are familiar with local cultural contexts. The third author of this study was responsible for collecting data and has been familiar with the relevant neighborhoods for more than 10 years.

\section{Results} 
We report our results by identifying what audience related concerns Hijra have on social media, explaining their connections with platform affordances and skill, and describing the contexts in which they shift across social media ecosystem. Exploring these results will provide us better understanding on Hijra's online self-presentation and participation.

Our results indicate that Hijra have complex gender and identity constructions. All of our participants mentioned their gender to be "hijra" in both interviews and FGDs. However, such construction of gender gets complicated when some of our participants mentioned classifying hijra further into \textit{Meye} hijra (for intersex) and \textit{Chele} hijra (for trans females), which is an internal gender classification some Hijra follow and not officially addressed by any legislation. Even though we did not have any participants who were proponents of this classification, it exists in some Hijra communities. Our data also suggests that despite mentioning gender to be "hijra", not all of our participants were happy with this legal term to explain their Hijra identity. As the term symbolizes historical exclusion of Hijra from the mainstream society, 13 out of 61 Hijra (around 21\%) participants in our study preferred to identify themselves as "Third Gender" than "hijra".  
Additionally, 19 out of 61 participants (around 31\%) mentioned their desire to be perceived as women although they do not fall under the traditional definition of cis-gender female. They experimented with their feminine identity through cross-dressing as traditional Bengali women while growing up (wearing saree, make up etc.). 
For a better understanding of our readers, we have created a chart (see \textbf{Fig. 3}) that represents how Hijra classify their identity in Bangladesh.
\begin{figure}
\centering
  \includegraphics[width=.4\columnwidth]{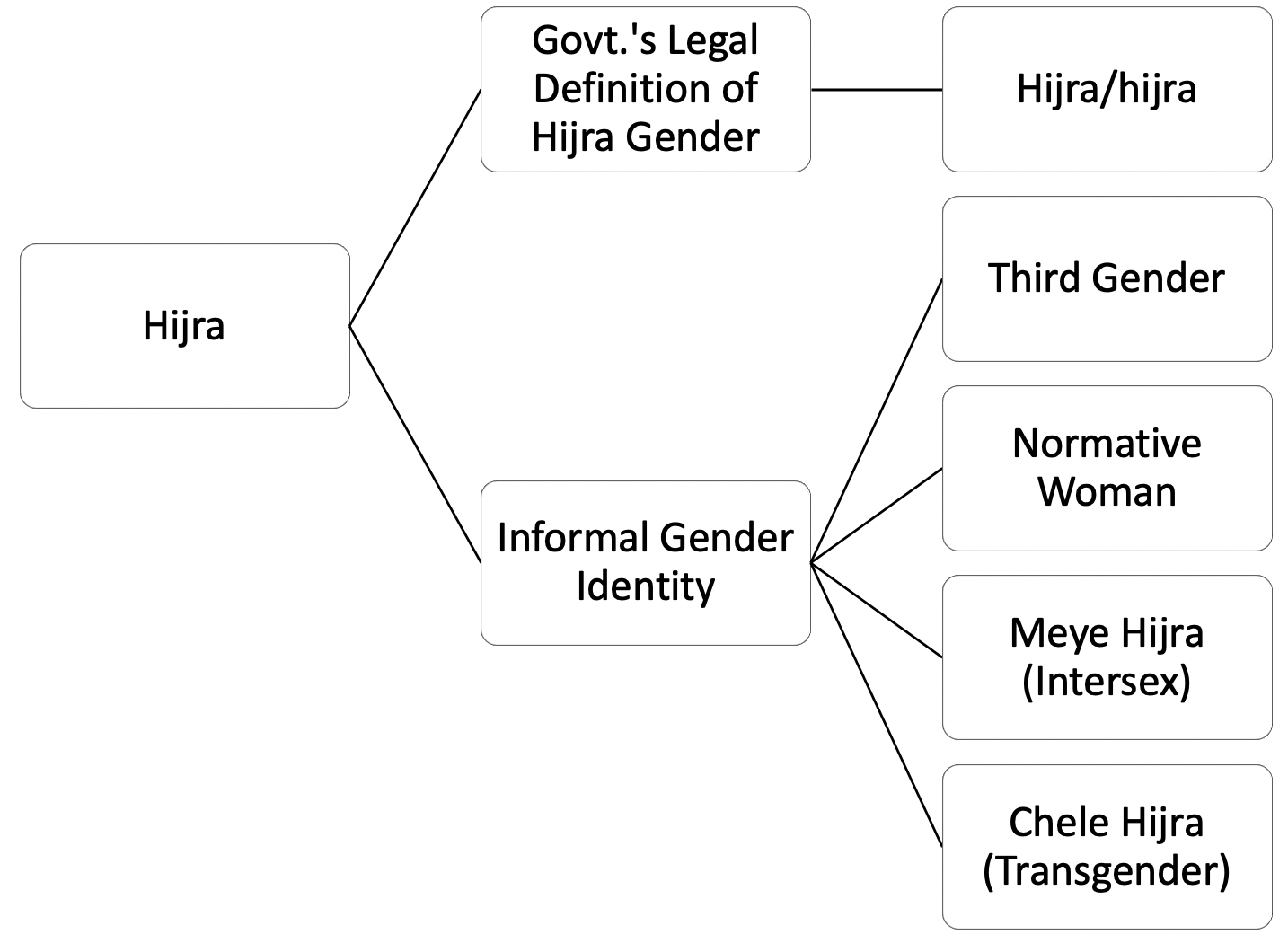}
  \caption{Hijra gender and identity classifications}~\label{fig:figure3}
  \vspace{-2em}
\end{figure}

Due to such complicated identity constructions, Hijra extensively employ their personal social media ecosystems to meet a number of different goals, as motivated by perceptions of platform audience, affordances, and spaces. However, our results also strongly indicate that local cultural and community influences, such as authority figures or group dynamics, as well as technical skill/knowledge also motivate the movement of activity and content across one's ecosystem, even sometimes overriding the importance of affordances. In this section, we will specifically focus on the intersection between Hijra's audience concerns and perceived affordances. Then, we will look into online participation shifting behaviors within Hijra to paint a holistic picture of gender minority self-presentation within one’s personal social media ecosystem in a non-Western context. 


\subsection{Audience Concerns and Uses of Social Media (RQ1)}
RQ1 asked what audience-related concerns Hijra consider when using social media, each reflecting a core reason for using social media. Our data suggest that Hijra in Bangladesh have three types of primary audiences on social media, with conflicting sets of disclosure concerns: family, other Hijra, and cisgender men. Uses ranged from daily communication to online sex work. Each of these audiences not only helped our participants form their personal social media ecosystem, but also shaped their online behaviors and concerns that help them to construct their identity online. 


\subsubsection{Family and Personal Connections}
Our data indicates that for 37 out of 61 participants (around 60\%), one of the primary audiences of interest are their existing family and friends. While family and friends are a common audience for social media content generally, the requirement that Hijra live away from their families in separate communities \cite{hossain2017paradox}, as well as persistent stigmatization by larger society \cite{aziz2019social}, heightens the importance of social media for reaching this audience. For example, interviewee X4(Age 18) mentioned:
\begin{quote}
    I live away from my family...we are not accepted in the society...and they (mass population) even leave the place if we sit beside them in public transportation...(which is why) it is easier for me to be connected with my family and friends in Facebook
\end{quote}
Social media platforms - and for Hijra, especially Facebook - have made it easier and more convenient for our participants remain connected with those they are close to while also respecting the structures of being a Hijra and avoiding public humiliation in the physical realm. However, this convenience does not come without complications. Despite the obvious utility of Facebook for Hijra, it can also create new anxieties related to heightened concerns over managing and selectively disclosing gender identity online. The need to be connected with family via Facebook can collide with the discomfort or serious disclosure concerns of the many Hijra, who then have to hide a portion of their online participation from their families. As interviewee X11(30) mentioned: 
\begin{quote}
    One of the biggest things in my life is that my family doesn’t know that I am Hijra... I have to do everything, specially in Facebook, by hiding my own identity...often it becomes very hard
\end{quote}
Similarly, participants like X8 and X10 also mentioned putting in extra effort to keep the platform they use for family connections walled off from the rest of their online life, such as strictly never using their \textit{meye nam} (female names as Hijra adopt as hijra) or not adding anyone from the Hijra community. As X8(25) mentioned mentioned,
\begin{quote}
   I use male as my gender (on social media)...(although) I love to think myself as a female. I have my family in my Facebook profile and they don't know about my identity...I like to keep it that way
\end{quote}
While such strategic outness online can maintain audience-related boundaries for Hijra, and therefore safeguard Hijra identity, participants still describe this as a "struggling" or "uncomfortable" position. For some Hijra, this concern over audience extends to a hard choice to completely hide their Hijra identity from their families as well as the rest of the world on platforms such as Facebook. Viewing Hijra identity as one of the major obstacles to use social media platforms without judgement, some Hijra prefer to suppress their own identity in order to enable their families to still be part of their online audience. This suppression of online identity, however, directly conflicts with crucial benefits of social media platforms. For example, it prevents receiving social and instrumental support that is tailored to one's actual, and not sanitized, identity and life experience, which past work with sexual and gender minority communities suggests is crucial to making said support effective and worthwhile \cite{brewster2013navigating, weisz2016out}.

\subsubsection{Hijra-Hijra Connections}
Hijra also use social media to seek guidance and suggestions from their fellow Hijra. 22 out of 61 (around 36\%) participants in our study said that connecting with other Hijra in this way, which enables finding appropriate communities in the physical world, is a crucial function of social media, as participant P38(40) noted in FGD4: 
\begin{quote}
    Now-a-days, we get connected with Hijra from the online communities we have in Facebook...In early days they physically had to find out the house of other Hijra...now due to the easy access to internet and Facebook pages, they can find us easily
\end{quote}
As Hijra are often forced to leave their home due to their identity, online communities such as "Badhan Hijra Sangha", "12 Vaja", and "Bangladesh Hijra Kalyan Foundation" (see \textbf{Fig. 4}) serve as a crucial tool for easing Hijras' transition from their home to Hijra communities. These online groups function as one of the primary tools for finding Hijra followers, Gurumas, and NGOs in the offline world.

This ability to connect via social media groups plays an outsized role in social and especially psychological support around the trauma many Hijra experience in their daily lives. Our participants were clear that their experiences include being regularly sexually harassed and potentially even raped entirely due to being perceived as vulnerable identity and lower status by the mainstream normative population, and there may be little or no support for combating or even just recovering from these incidents in their own home communities. This vulnerability and the lack of support around it make online platforms crucial spaces for sharing extreme harassment experiences which originate offline or even in other online social spaces, as sharing these experiences can be crucial to dealing with the psychological fallout and reducing dangerous feelings of isolation. Additionally, online spaces for Hijra are often the ideal starting point for receiving tangible, instrumental support, as while they provide Hijra in distress a chance to be heard by other Hijra, they also connect Hijra directly to the Gurumas who run these online spaces, and may have resources to share. Participant P14(38), who is a Guruma, from FGD3 mentioned,
\begin{quote}
    I took in Ruma (pseudonym) at the age of 7. Ruma was being sexually assaulted by more than 30 men from Ruma's local area just because Ruma was a hijra...I got to know about Ruma from Facebook as Ruma shared those experiences in our online community looking for help...I took help of an organization and rescued Ruma to a safe place
\end{quote}
For many like Ruma, groups have been an escape and a crucial resource for many Hijra who were rescued by Gurumas or NGOs who work with sexually abused Hijra. The groups provide a bounded space and a distinct audience for Hijra, and especially provide an opportunity to be open about traumatic events without being entirely public. However, notably, the need for these kinds of support, services, and connections sometimes drives those that are not in groups, or who fail to find the support they need in groups, to post their experiences more broadly Instead of sharing in any group or community pages, they share those experiences in their personal social media profiles, hoping to be seen and heard by other Hijra. For example, during FGD3, participant P26(20) said:
\begin{quote}
Well, we found her posting on being sexually abused that she shared in her profile...it came to our attention through multiple sharing...and our Guru Ma wanted to save her and now she’s here with us (in the hijra community)
\end{quote}
Pursuing the specific audience of Hijra, and finding the appropriate platform within their personal ecosystem in which to find Hijra who might be in need, allowed P26 to find and help Hijra that are so isolated as to never join an online Hijra group. 

Through social media groups, and individual connections, Hijra employ their personal social media ecosystem to find the right audiences from which to draw support, so they no longer have to suffer alone through traumatic, abusive experiences. For most, social media groups provide a place to share and be supported around experiences that they could not share physically or virtually through a sanitized profile or a general-audience social media group/page. Moreover, by allowing connection and solidarity between Hijra, platforms with Hijra audiences can allow Hijra a space to not hide their online identity, but rather receive direct, psychologically-beneficial support from their peers by fully expressing it - both the good and the bad.



\begin{figure}
\centering
  \includegraphics[width=.8\columnwidth]{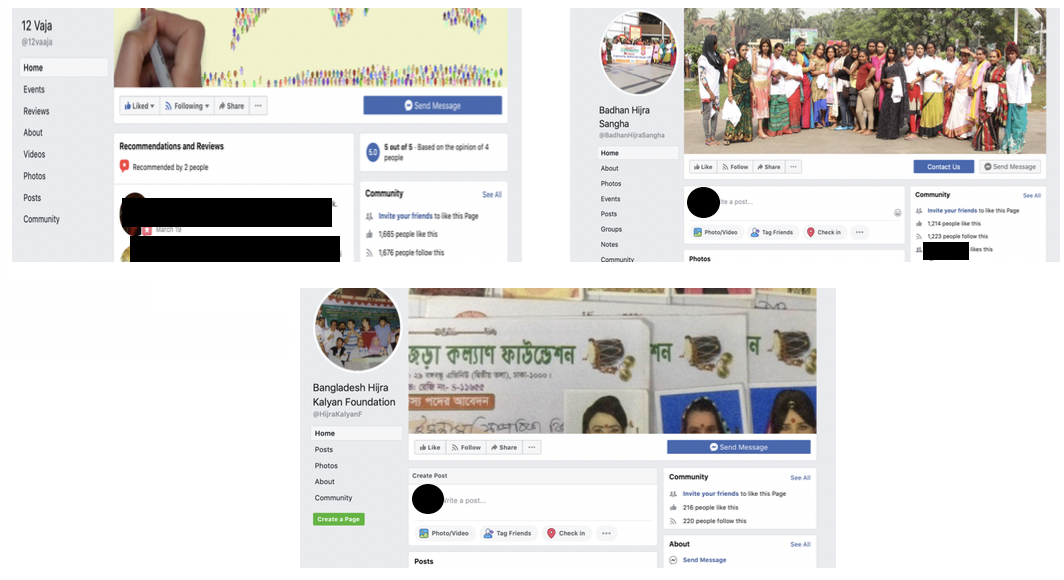}
  \caption{Different Facebook Online Communities of Hijra in Bangladesh}~\label{fig:figure4}
  \vspace{-2em}
\end{figure}
 
 \subsubsection{Cisgender Men Connections} The final, and most potentially fraught, audience Hijra pursue is cisgender men. 36 out of 61 participants (around 59\%) mentioned being purposefully connected with cisgender men (individuals who identify as men and were assigned male at birth) through different social media platforms with an expectation of developing a romantic relationship with them. In FGD3, participant P18(18) said:
\begin{quote}
    Hijra people mainly starve for a guy’s company...We add them (online)...we hang out and have fun online all the time
\end{quote}
Hijra largely turn to social media, as it provides them a convenient platform to access cisgender men. Building relationships with Hijra is prohibited under sociocultural, religious and political rules and customs \cite{khan2009living}, which potentially exclude them from having any sort of traditional relationships with men. Hence, social media becomes a crucial channel - potentially, the only viable channel for many Hijra, as it allows them an opportunity to have a romantic or even just flirty relationship with men, and to explore basic relationship possibilities. 

Interestingly, this audience is paired with an audience many Hijra explicitly avoid online: cisgender women. As interviewee X4(18) said:
\begin{quote}
    We add [men] online who are really handsome...we love talking to them...If we find them not pretty enough we remove them... But we don’t add any women [cis-females]. We envy them. We have a fear men will stop hanging out with us, if we add them online
\end{quote}
Our participants say they often feel overwhelmed by having cisgender women (individuals who identify as women and were assigned female at birth) on the same social media platforms they frequently use. Cis-women are seen as something of a threat to Hijra when men are also on the platform, as in that context they are viewed as stiff competition for the attention of men. As interviewee X4(18) said:
\begin{quote}
    Here’s the difference between a Hijra and a woman...A girl can hardly be with a man in this society but as a Hijra I can sleep with ten other men at the same time. It doesn’t bother me neither the society...Sometimes we feel so happy that we think a woman doesn’t even get the pleasure of being (sexually) with a man like us...But then again, we get sad thinking no man we meet online will ever marry us just because we are Hijra
\end{quote}
Due to their uncomplicated gender identity and sexuality, cis-women serve as a constant reminder to Hijra about their comparatively socially excluded and disadvantageous status. In this way, the presence of cis-women as an audience on some platforms complicates audience management should desirable men be on the same platforms, especially in situations where Hijra are essentially perceived as sex objects. Being desperate to be accepted by normative cisgender men on online platforms, where women are avoidable, influences many Hijra to exclude cisgender women as audiences in their social media ecosystem.

\begin{figure}
\centering
  \includegraphics[width=.7\columnwidth]{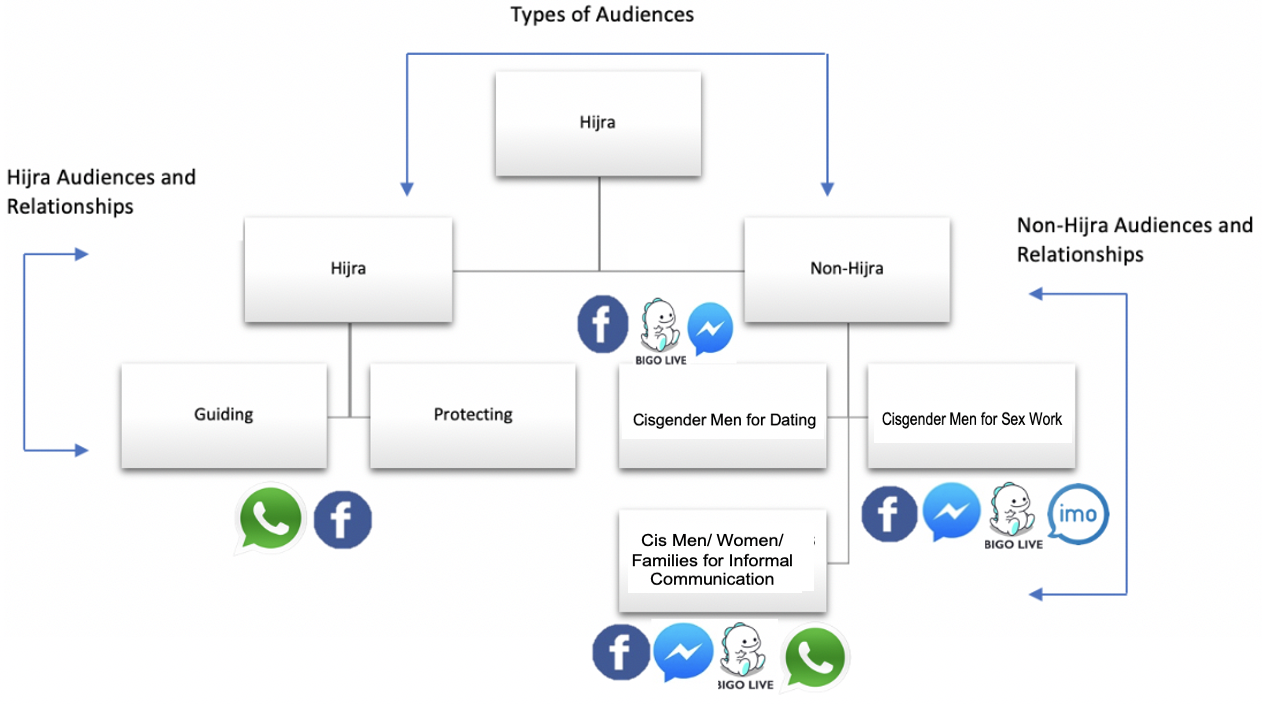}
  \caption{Types of Audiences and Relationships of Hijra in Distinct Social Media Platforms}~\label{fig:figure5}
  \vspace{-2em}
\end{figure}
Hijra also connect with cisgender men on social media platforms to find opportunities to generate income via sex work. As it is hard for Hijra to land traditional jobs due to their gender nonconformity and overall social status \cite{aziz2019social}, earning money through sex work often becomes a primary livelihood. As participant P19(40) said during FGD3, social media is often the most accessible way to set up this sex work:
\begin{quote}
    Here (in Hijra community) people can be illiterate, but they surely can use Bigo Live, Facebook and other audio/video applications like IMO, Messengers... Because, they can earn money without any toil
\end{quote}
This ability to find sex work is a primary motivation for many Hijra for both using social media and expanding their networks to include cis men. Some participants even mentioned being connected with clients from otherwise-inaccessible foreign countries (such as Pakistan, India, Saudi Arabia etc.) for sex work through these platforms. Interviewee X5(20) mentioned:
\begin{quote}
    Three of my clients from Bigo Live and Facebook actually came to Bangladesh from abroad to meet me...One of them offered me to go with him and live a married life in abroad
\end{quote}
Having potential clients as part of their social media audience makes it easier for many Hijra to not only make a living via sex work, but also expand their opportunities via gaining popularity beyond physical borders. While these goals and uses do not align with the core aims of platforms themselves such as Facebook or Bigo Live, they nonetheless remain crucial for many Hijra. However, such functionality of using social media platforms to earn money through sex work did not come without any consequences for many Hijra. 
While disclosing their identity online provides Hijra opportunity to earn money, it also puts them at risk of compromising their privacy and hijra identity online. Despite of being aware of such disadvantages, many Hijra are forced to decide between their privacy and online identity disclosure to ensure their livelihood through sex work.

Audiences are an integral part of Hijras social media ecosystem, playing a large motivating role in terms of both identity/disclosure management across platforms as well as achieving particular goals of social meia use. This then plays a large role in setting Hijra's overall expectations from online participation (\textbf{Fig. 5}). Whereas certain audiences are expected and welcomed in the ecosystem, same and other audiences can cause discomfort and unwanted existential crisis for some Hijra - a problem often resolved by spreading audiences across one's personal social media ecosystem.



\subsection{Affordances of Social Media Platforms for Hijra (RQ2)}
As we have demonstrated above, Hijra must balance multiple audiences with very different orientations towards Hijra identity, and therefore conflicting disclosure requirements, in order to derive both social and instrumental benefits from social media use. As prior work has shown, available affordances, especially for control over audiences and visibility of content, have a large impact on both decision-making around self-presentation and identity disclosure \cite{devito2017platforms} as well as platform choice within a social media ecosystem \cite{devito2018too}. This holds true for Hijra, who must carefully consider their identity management options and the tools available to them. 

Affordances (possibilities for action/features that each platform offers) to users play a big role in how users perceive and experience distinct social media platforms, often for distinct audiences. Our data suggest that, as Hijras social media ecosystems consist of different platforms \textbf{(see Fig. 6)}, afffordances offered by each platform have an influence on their preferred online participation. Distinct social media platforms offer distinct means to control audiences online, a primary concern for Hijra. According to our data, the most popular social media platforms for Hijra were Bigo Live, Facebook, Messenger, IMO, and WhatsApp. Many participants used these platforms almost daily to maintain their communication, audiences and sex work.

\begin{figure}
\centering
  \includegraphics[width=.8\columnwidth]{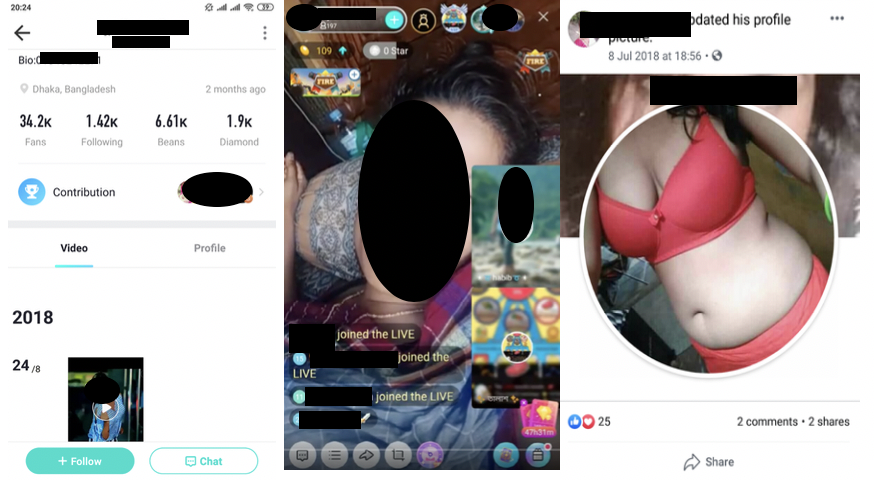}
  \caption{Different Social Media Platforms Used by Hijra (images are collected and shared with participants' consent)}~\label{fig:figure6}
  \vspace{-2em}
\end{figure}

\subsubsection{Presentation Flexibility}
Our participants recounted how certain suites of affordances help fulfill their individual audience management needs. For example, Hijra primarily maintain communication with their families and close friends using Facebook.  
Our participants pointed out several affordances that make Facebook easy to use when it comes to communicating with family on a daily basis. While our participants recounted using multiple platforms with family, Facebook was seen as the preferred platform overall due to the affordances provided. For example, compared to other platforms, Facebook affords more of what DeVito et al. call "presentation flexibility" \cite{devito2017platforms} or the ability to use multiple formats and styles to present oneself to others. For example, interviewee X15(20) said, 
\begin{quote}
    I post my photo, chat with friends and families in there (Facebook)...Facebook also helps me not only to share my inner feelings with them through sharing posts but also let me do check-ins to let them know where I am or where I am going...it is so easier to update them about my life...I can also do audio or video call in Facebook Messenger 
\end{quote}
Our participants expressed a desire to use robust media types, not just text and photos, in maintaining these family relationships, and also noted the importance of having all these media types available on a single platform so as to not fragment the family audience. 

However, for some Hijra (13 out of 61, around 21\%), Facebook has not always been an ideal platform to utilize this presentation flexibility. While multi-factor media sharing within Facebook has provided Hijra robust ways to express themselves to their family connections, it has also potentially restricted their flexibility to disclose their true identity in front of these audiences. Even though disclosure of hijra identity was a sensitive information to our participants (such as X8, X10), some Hijra preferred to share this information on social media willingly with their family connections as a way to express their true self online; however, they were unable to do so. According to interviewee X13(24),
\begin{quote}
    No, I did not give hijra as my gender online\dots Even though my family knows about my hijra identity and I am openly Hijra in Facebook, I have put female as my gender as an alternative option\dots I know I am not a female but actually in our country there are only two options of gender in Facebook- male and female\dots no where Hijra option is given 
\end{quote}
Hijra's presentation flexibility on social media is significantly effected by the affordances Facebook has provided to its users. Despite Facebook's continuous effort to provide its users flexibility expressing their preferred identity online, it particularly fails to assist Hijra- who wish to share their hijra identity with their audiences online, including family connections. Facebook's "custom" option to specify own gender, apart from male/female, by the users themselves does not work for Hijra, as (our participants mentioned) the feature is confusing and unfamiliar to them. For example, from FGD5 participant P43(35) mentioned,
\begin{quote}
    We can choose hijra (as gender)? How? I can only see male and female in the options\dots There was a third option probably ("custom"), but I had no idea what that meant. Does that mean hijra?
\end{quote} 
Term such as "custom" is unfamiliar within Hijra, and it provides no information to them regarding their flexibility to choose their own gender in their preferred online platforms. 
As Hijra is not explicitly included in the gender spectrum of social media platforms, it potentially restricts and forces the members of Hijra community to construct their identity online within the dichotomy of male and female.



As our data suggests, while Facebook is a big part of Hijra connection with their families, by way of contrast, in flirty communications or sex work situations, it is more useful for Hijra to have access to a platform with far more limited presentation flexibility, so as to bound the possible interactions they can be expected to have with potential audiences or clients. For example, 28 out of 61 Hijra participants (around 46\%) preferred to use IMO while building romantic rapport with cis-males due to its primary focus on chat and limited features on presentations formats. From FGD1, participant P1(18) said, 
\begin{quote}
    Through IMO, I talk to the men who are nice and interesting...I usually do audio calls through IMO and use my female voice to present myself as woman...they never realize that I am a hijra
\end{quote}
IMO's core functionality of regulated one-to-one or group conversations through only audio, video, and written chat (unlike Facebook that also includes other features within the platform) helps Hijra to represent themselves in a way that benefits their goal of building a romantic rapport. Some Hijra also mentioned using IMO and its limited features and affordances to engage in sex work that is carefully curated only for their clients. Regarding this, interviewee X10(35) mentioned,
\begin{quote}
 I am engaged in sex work through IMO...I like to use it...it is simple and easy and does the work for me...I do video chat there and collect money afterwards
\end{quote}
As chat, and not additional functionality such as games or item listings, takes center stage on IMO, the tightly-specified form of limited presentation flexibility afforded on this platform makes it easier for Hijra keep interactions bounded to sex work and nothing else. 

\begin{figure}
\centering
  \includegraphics[width=.6\columnwidth]{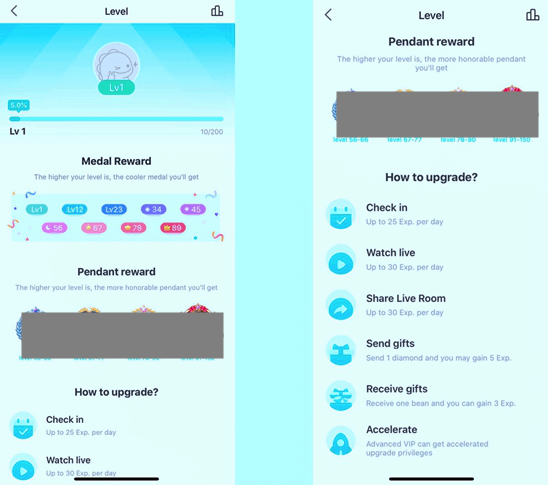}
  \caption{Leveling up in Bigo Live}~\label{fig:figure8}
  \vspace{-2em}
\end{figure}
\subsubsection{Audience Transparency}
Privacy concerns are, of course, paramount for Hijra when making decisions as to how to fulfill their needs via social media. One of these concerns relates to what DeVito et al. call "audience transparency," or the afforded ability to be aware of who is in one's audience. Affording increased awareness of not just who is in one's audience, but also the relevant characteristics of those audience members, has proved useful to Hijra pursuing sex work in particular, as interviewee X11(30) said,
\begin{quote}
    Actually I don’t face that problem (privacy issue) in IMO... By making a call through IMO, I can understand how old are they... In IMO, who have my phone number only those people can contact with me. That’s why I like to use IMO more than Facebook
\end{quote}
Being engaged in a profession like sex work, concerns regarding who they connect with becomes an important aspect for many Hijra online who wish to stay away from being harassed or bullied from unwanted prospective clients or others. As it is relatively easy to find someone on Facebook even without knowing their full name or information, it becomes important for many Hijra to hide their personal profile or identity from certain audiences, who can be their potential harassers in future. As many Hijra's livelihood is dependent on online sex work, they prefer to keep it as safe as possible using platforms like IMO, where clients or audiences with Hijra's personal phone number can only reach to them. 
However, exposure to a broader but targeted sex work client base has also been an important part of Hijra sex work online that conforms audience awareness. Some Hijra (15 out of 61, around 24\%) preferred to use Bigo Live for its feature of level that helps users to broadcast their live video to a wider audience, such as foreign sex clients, with similar interests. For example interviewee X2(35) mentioned,
\begin{quote}
    Having higher levels in Bigo Live helped me to expand my fan base and to connect more with foreigners...the higher the level, the more possibilities that people and foreigners will see my broadcasts and videos...for the kind of work I do (sex work), I prefer to be connected with clients who are foreigners...They not only pay more but most them are also nicer and more polite
\end{quote}
Bigo Live allows users to live-stream their favorite moments, and make friends from all around the world through live video/audio/text chat \cite{bigo_live_stream}. It also offers its users to boost/promote their contents to compatible users through increasing their level within the app. One can level up in Bigo Live by logging in every day, sending virtual gifts to other users and so on (\textbf{see Fig. 7}). To make it simple, the higher the level, the more popular one is and the more easier it is to earn money.  Through engaging more on the platform and increasing the level in Bigo Live, some Hijra expect to be seen and discovered by foreign users who are interested in sex work and will provide big amount of money for their work. Distinct feature offered by Bigo Live has allowed Hijra to expand their popularity beyond the border and created an opportunity to be connected with expected audiences. Even though for some such exposure may be a privacy issue, for many Hijra it offers a strategic process of selecting clients or audiences to improve their professional lives.


Such preference of relationships and audiences in distinct social media platforms provide us some significant insights on the perceptions of Hijra in Bangladesh who become a part of distinct social media ecosystems. 


\subsubsection{Visibility Control and Identity Persistence}
Although distinct social media platforms have provided Hijra ways to communicate or earn money through sex work, they also have brought additional harassment, making the platform's afforded level of visibility control, which we refer as presenting themselves with selective visibility, essential concern when countering harassers. For example, many Hijra take advantage of the block feature in Facebook when the level of derogatory language used against them on their own social media platforms becomes intolerable. During FGD1, participant P1(18) mentioned:
\begin{quote}
    For example, in Facebook or Bigo Live, when I upload a picture or video, people make comments like “Hijra” “hot/sexy”, “show me your naked body” etc. I instantly block them from there
\end{quote}
The blocking features on Facebook and Bigo Live help Hijra maintain their social media profiles by preventing unwanted harassment. However, in some cases, the blocking feature alone is not enough to afford enough visibility control to Hijra to adequately combat harassment and bullying. Specifically, our participants report that this is a problem with specific harassers who create multiple account to circumvent blocks. On this issue, interviewee X2(35) said,
\begin{quote}
    ...some people just keep calling and harass me online... when I cannot tolerate any more, I block them...The  irony is some of them open new accounts and add again...one day I may find out that is the same guy that I blocked
\end{quote}
Here, we see a conflict between a platform's afforded high visibility control (individual, fine-grained block tools) and afforded low identity persistence (easier alternatives to the tools) in terms of audience management for Hijra. While the platforms are, indeed, trying to afford better visibility control to ensure safety, there they are sometimes simultaneously providing ways to create new/multiple accounts online for its users that bypasses the usability of the blocking feature in the first place for Hijra.

\begin{figure}
\centering
  \includegraphics[width=.6\columnwidth]{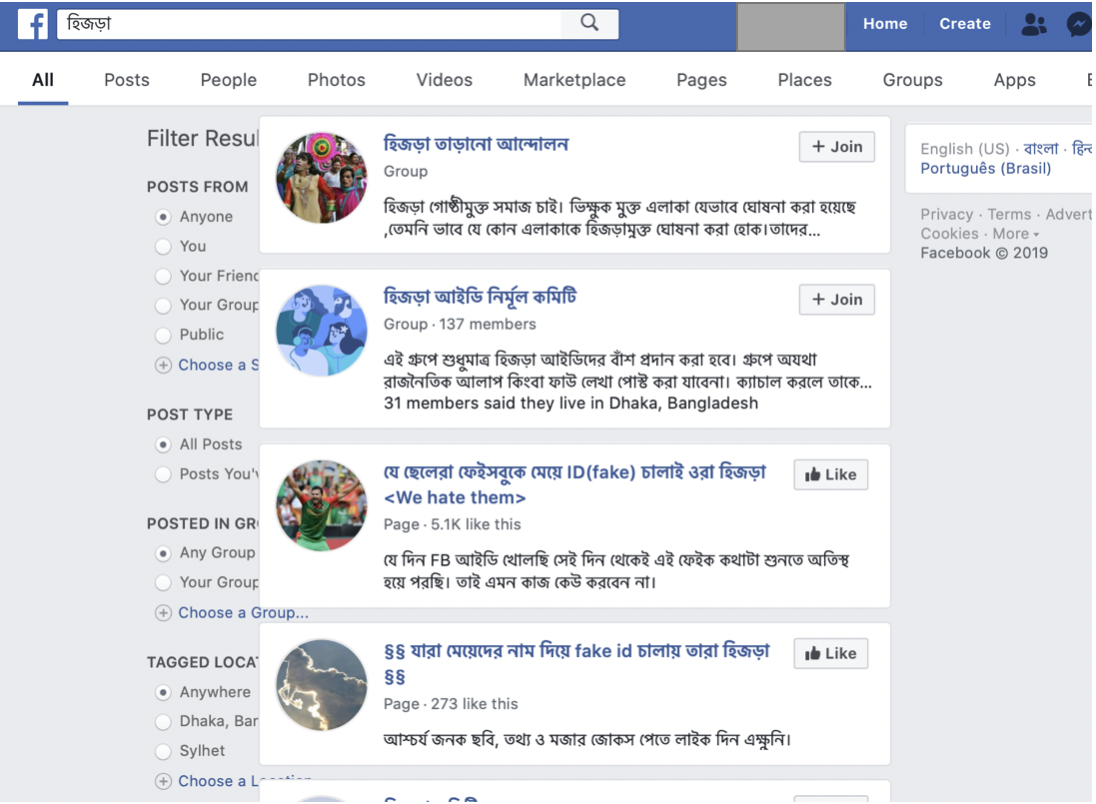}
  \caption{Anti-Hijra Groups in Facebook}~\label{fig:figure9}
  \vspace{-2em}
\end{figure}
Inability to control who gets access to sensitive information and how 
has been a big issue for Hijra as it often turns into a matter of serious privacy and security concerns. Leaking personal photos/videos publicly, using personal information to create fake profiles and seek money from others and spreading rumors have been common events for Hijra who were either unable to restrict the harassers or their contents. 
On this topic, interviewee X6(25) said, 
\begin{quote}
    ...he was sharing personal photos of me and was asking for money from others in Facebook...I could guess who it was but was not sure...one of my colleagues suggested me to disable my id and report the culprit to prevent him from accessing my personal information...but the problem was I had no idea how to disable my ID or to report to Facebook about this event and the culprit
\end{quote}
Many Hijra lack the necessary training and technical skills to effectively employ platform tools, even to the extent that it seriously impacts their security online. While a majority of the users may be familiar with features like blocking someone or deactivating personal id to ensure visibility control online, for some Hijra, these privacy tools may not seem easily accessible, and visible, but rather excessively intricate and complicated. Being unable to utilize such complicated privacy controls has also been a concern for our participants while reporting different anti-Hijra online communities that promote hate speech against Hijra \textbf{(see Fig. 8)}. Even though platforms like Facebook provide users security tools to report online hate speech and harassment, Hijra are seldom made aware of those tools to protect their content or shape their audiences, as participant P5(20) demonstrated in FGD1:
\begin{quote}
    Can we report these online communities in Facebook? Really? How?... What is a \textit{report} option? I am sorry...We are not good with technology... we don't know how to use this option online
\end{quote}
To our participants, tools to  
report online community pages are often unfamiliar, as there is no clear instruction or indication provided to them regarding their options to file complaints against harassers. Even though both block and report are privacy tools to control unwanted events and individuals online, Hijra are not aware of all these options due to their limited knowledge on the platform's affordances. 
Not being familiar with options like report, and not seeing any visible action against such harassing pages, individuals or contents have forced many Hijra to perceive platforms like Facebook as more hostile towards them. Such limited technical skill-set and knowledge/awareness at controlling their contents, events and individuals intensify Hijra's vulnerability online and make their social media ecosystem more complicated.

 \subsection{Shifting of Participation Across Social Media Ecosystem (RQ3)}
Unexpected negative events and harassment can trigger Hijra to shift, limit or stop their online participation through different social media platforms. As we explore RQ3, our data suggest that Hijra adopt strategic decisions to shape their participation online that are often motivated by the platforms' affordances, Hijra skill-set and community influences conforming those negative experiences.  

\subsubsection{Migration, Limited Participation and Social Media Non-use}
Hijra's experiences of being severely bullied and harassed online give Hijra good reason to identify which online spaces, which DeVito et al. define as the user's own conflation of platform (including affordances) and audience \cite{devito2018too}, are less likely to be unsafe. Even though Hijra face extreme harassment online, instead of deleting their profiles from distinct social media platforms, 11 out of 61 Hijra (around 18\%) prefer to shift from one platform to another in a hope to search for a space that will provide better experiences in terms of their identity. As interviewee X9(18) mentioned:
\begin{quote}
    What I feel best about Bigo Live is it has less harassment...Facebook promotes more harassment...which is why I left Facebook and moved to Bigo Live...I still have my account in Facebook but I don't use it anymore 
\end{quote}
As mentioned in the previous section, Facebook is perceived more hostile towards Hijra, Hijra looks for spaces that are supportive towards them and their identity. 
However, it is not uncommon for them to keep their old profiles open. Such a decision to migrate from Facebook to other social media platforms comes with it own consequences. As Facebook is the primary media for many Hijra to be connected with their loved ones, including friends and families, such migration puts a dent in their virtual social lives forcing them to compromise their participation online. Whereas on one side, Facebook is working as an alternate for social interactions for many Hijra, stressors like additional negative experiences and privacy concerns force them to transfer or migrate their participation elsewhere. 

Apart from migrating their participation, many of the Hijra (29 out of 61, around 47\%) mentioned either limiting or withdrawing their participation from distinct social media platforms to protect or save their identity as Hijra online. For example, interviewee X13(24) mentioned,
\begin{quote}
   There are many boys who come in live in Bigo Live, disguise themselves as girls and make vulgar and defaming videos...There is no way to distinguish the fake Hijra from us in Bigo Live...For them other people blame and shame Hijra like us...For these reason, I don’t go in live or use Bigo Live much now
\end{quote}
Social media platforms like Bigo Live has no tool to identify fake Hijra or fake profiles of Hijra that often try to defame the whole Hijra community online. For such cases, it becomes hard for many Hijra to participate online freely, which often leads to limited participation in the space or platform. Apart from this limited participation, our data also suggested social media non-use within our participants. Some Hijra mentioned withdrawing their participation from distinct social media platforms because of getting hacked and being unable to retrieve the profile. As X12(29) said:
\begin{quote}
     My Facebook Messenger got hacked...the person who hacked it asked for money to others and leaked some of my personal photos...I did not know what to do or how to stop it, so I stopped using it [Facebook Messenger]
\end{quote}
Limited skills to handle concerning situations like profile being hacked has also forced Hijra to withdraw their participation online. Even though our participants were familiar with the use of social media platforms like Facebook Messenger, they were not familiar enough to handle situations like these where a little bit more knowledge or skill on using the platforms was required. Such experiences were concerning for Hijra, which lead them to stop and leave the platform for good costing their proper interaction their families or clients through those platforms.
\subsubsection{Community influences}
The decision to shift to another platform in one's ecosystem also sometimes depends on the group dynamics Hijra value within their communities. As Hijra have strong bonds within their community, their decisions to move platforms or use certain platforms in certain ways often get influenced by what others from the community suggest, or the information peers or authority figures provide. Participant X6(25) (a Guruma from Comilla) mentioned:
\begin{quote}
   I heard if you do drugs or something bad, your online ID will be hacked. But BIGO Live is safe though...I also suggest my fellow Hijra to be safe when using these online platforms
\end{quote}
Often, along with fellow Hijra, authoritative figures such as the Gurumas have power over what a follower should know about or use for their online participation. Even though the information provided by the Gurumas are not entirely correct or true every time, due to the influence they have over their followers, it effects the way other Hijra shape their participation in the ecosystem. The strong bonds within Hijra communities, which are crucial to internal solidarity and mutual protection, provide reasons for Hijra to shift their participation from one platform to another based entirely on this type of personal say-so. This collective mentality is also sometimes visible within Hijra online in a larger sense. Some of our participants also mentioned a tendency to adopt or reject distinct social media platforms based on their collective experiences or own internal group dynamics. 
Related to such group dynamics, from FGD1 participant P1(18) said,
\begin{quote}
    Well, we use smartphones. We eventually know about different social media platforms...If someone in our community uses or prefer any specific one [social media platform] for you know...different reasons, we all get to know about it and try it out
\end{quote}
When a member of the community introduces a new app or platform that is beneficial (in terms of privacy, better communication, clients etc.) for the whole group, it often influenced our participants to shift their social media participation to that new platform. As solidarity within Hijra communities is strong, it guides Hijra to adopt a new technology or social media platform. 

While existence on social media becomes an issue of safety and privacy for Hijra, restricting participation online seemed more feasible for many of our participants. By limiting, withdrawing and shifting participation from one social media to other, Hijra strategically try to control such instances as much as possible with a cost of their smooth participation online.

\section{Discussion}
Our study highlights important aspects of social media ecosystem of Hijra, a GSM population who have their own specific audience concerns and perceptions of affordances on different social media platforms. While RQ1 and RQ2 specifically inquire into Hijra's social media ecosystem to understand with whom, how and where they build their connections and perceived affordances, RQ3 digs further to see where the system fails and thus force Hijra to migrate or shift their participation on online social media. 

\subsection{Gender Minorities, Hijra and Social Media Platforms} 
As our results suggest, Hijra break through traditional gender boundaries and cannot be reduced to merely metonymic, Western figures for an analysis of gender fluidity 
\cite{knight2014surveying}. During the study, it was interesting to explore participants' choice of terminologies to define their identity that do not fit under the existing English language terms for other GSM communities. It is possible that many among the LGBTQ+ community from Western context are unaware of most of these local terms used to define GSM identities. In Western contexts, many LGBTQ+ communities use the term queer as an "umbrella" term that encompasses all who identify as lesbian, gay, bisexual, or transgender \cite{zosky2016s}. However, due to Hijra's unique multi-classifications of gender perceptions (that may not conform to the normative social expectations and definition of gender minority and sexuality), it is unfair to try and explain Hijra identity using English LGBTQ+ terminology \cite{ibahmed_2017}. Whereas non-Western mainstream societies fail to articulate the concept of "Meye" (for intersex) and "Chele" (for transgender) within Hijra communities by forcing "hijra" as general gender term for all Hijra, there 
it is even potentially offensive to use Westernized gender and sexuality categories for them. Even in the Western
world, the categories that are used to define GSM are not self-evident, and raise the need of asking localized questions
on what these categories mean to the people in a specific
country \cite{lubbe2013lgbt}. Patil \cite{patil2018heterosexual} describes this notion as a heterosexual matrix within predominant western epistemic frame that defines gender intelligibility from male-female dichotomy and poses a challenges for the people who can not be classified into a  normative or even expanded western schema. As a consequence, this study adds to the conversation on GSM that exists in non-Western context by exploring Hijra's identity perceptions, of which some are impossible to be translated or to fit into the Western models of gender and sexual identities \cite{milanovic2017impact}.

As a stigmatized gender minority group, Hijra lack social contact with others and therefore, online spaces like social media platforms hold particular value for this community people. While previous work on GSM \cite{blackwell2016lgbt, haimson2016transgender} discuss the struggles of LGBTQ+ populations being a part of the online spaces, our paper extends the definition by accounting Hijra from non-Western contexts into the conversations. The study shows that Hijra use different social media platforms to carry out certain sets of actions and expectations that are carefully distributed throughout their online participation. 
Whereas, the dominant mainstream population often considers the decision of being a part of an online space as a choice rather than a necessity for many stigmatized populations, Hijra's dependencies on social media platforms to achieve primary livelihood through online sex work or social life (that they are deprived of in the physical world) reveal their vulnerabilities and necessary dependencies on online spaces. Previous studies \cite{altaf2012comparing, chakrapani2004hijras, abdullah2012social} have explored such engagement of Hijra in the field of health or social science, but it is absent in the field of CSCW and social computing. Due to their complex identity, Hijra fall into the conundrum of choosing between online exposure and identity protection that in turn motivates them to look beyond only audience management objectives while navigating through different online platforms. This study highlights those complexities by engaging in deeper explorations on GSM social media ecosystems and evaluating the framework more intensely from non-Western context.

\subsection{Hijra Audiences, Affordances and Spaces}
Similar to the existing framework of social media ecosystem \cite{devito2018too}, our participants described relying on perceived audiences who are an integral part of their personal social media ecosystem. Audiences such as "targeted imagined audience” based on communal ties \cite{litt2016imagined} (close families and friends as well as other Hijra members of the community), or “outright” targeted audience (such as cis-males) \cite{devito2018too} play an important role in Hijra's online ecosystem that is constructed via interplay between spaces and affordances. 
Due to the stigmatization Hijra face for their identity, they are in constant search of audience awareness, controlled exposure and inclusiveness by segregating identity related contents in distinct spaces. For example, even though participant X11 strategically restricts sharing any hijra related content in Facebook due to having social and cultural obstacles (social image towards family and friends), X11 does not hold back disclosing those contents in Bigo Live or IMO because of the platforms' afforded safety and controlled exposure. For many Hijra, having tightly-specified form of communication space with easier, less complicated features heightens their presentation flexibility towards their audiences and builds a sense of control on who they connect with and how (such as participant X2). 
Often, affordances in Hijra's social media ecosystem are not enough to address their audience related concerns on self-presentations and visibility control. Even though the current lens of social media ecosystem focuses on the combination of audience and affordances in audience management, for Hijra, such interplay does not always work due to their lack of understanding of the platform's affordances as well as required technical skill-set and knowledge to manage audiences. 

According to DeVito et. al. \cite{devito2018too}, GSM populations from Western context tailor their online LGBTQ+ presentation via affordances they see as helpful; however for Hijra, same affordances become challenging and limited due to the perceived low identity persistence of the spaces and their limited technical skill set and knowledge (such as participant X6). Unable to control their self-presentation using provided privacy tools, many Hijra face unavoidable harassing experiences online, which in turn, impact the way Hijra perceive their audience management strategies. While total around 65\% of our participants strategically migrated or limited their online participation from one platform to another due to these negative experiences, they kept their personal profiles open and purposefully not managed in terms of audiences. Even though Western GSM populations have a tendency to follow rigorous processes 
in terms of sharing content with specific audiences online \cite{haimson2015disclosure}, Hijra (e.g. participant X9) tend to less so while shifting participation for audience specific reasons 
without erasing or fully closing previous profile at all, which directs our observations to the likelihood of their reversion \cite{baumer2015missing}. 
Due to Hijra's professional and personal objectives, they often need to ensure maximum exposure online; this may influence them to revert back to online spaces they left earlier. Current models of social media ecosystem needs to address such dynamic shifting of participation by GSM users who purposefully keep their audience management strategies relaxed while leaving certain social media platform. 

\subsection{Extending Social Media Ecosystem Framework}
While our data strongly suggest that Hijra are at many times aligning their personal social media ecosystems through Personal Social Media Ecosystem framework \cite{devito2018too}, it also clearly reveals areas where this lens must be extended to better account for non-Western contexts. Our findings on Hijra’s struggle with self-presentation on social media extends the existing lens of social media ecosystem by considering technical skill and knowledge as a fourth element in the framework. 
These findings align with the current literature on digital literacy that recognize user skill as an vital part of privacy control and must include GSM communities from non-Western contexts.

\subsubsection{A fourth element: Skill}
In our study, we have observed a strong influence of technical knowledge/literacy or skill on Hijra's way of addressing their audience concerns and spaces. This skill/knowledge represents additional factor that potentially and significantly shapes how stigmatized GSM populations outside the US construct their concept of audience privacy and spaces and direct their participation online accordingly \cite{nova2020understanding}. Past literature like \cite{devito2018too, devito2017platforms} have explored LGBTQ+ and their audience concerns through the online spaces and affordances but did not account the concept of skill that can highly impact the spectrum of social media ecosystem for Hijra. By zooming out from the current social media ecosystem framework and looking through a macroscopic lens, we can explore the comprehensibility of the affordances and users' specific skill sets that are required by the platforms to fully utilize the affordances. A prior study finds that technology design for marginalized people in mind are supposed to have low barriers of access \cite{garzotto2008marginalized}. Having tons of privacy tools for the users is not enough if they are not accessible, especially for the marginalized GSM populations like Hijra. Here accessibility does not mean access to the technology, but rather denotes the concept of accessing the knowledge that is required to reap full benefits from the platforms' affordances. Our data suggests that this struggle with technological understanding and skill is not an individual problem, but rather a common experience within many Hijra communities. \citet{sambasivan2010intermediated} have mentioned how in developing countries this type of struggle is not uncommon, as many people from low-income communities lack textual and digital literacy that effect their technology- operation skills. 
Hence, while we try to explore how stigmatized users from Western context interact with social media to construct their identity by only considering provided platforms' affordances, we may potentially exclude other stigmatized GSM communities from non-Western context whose perceived platform affordances are significantly impacted by their technical knowledge and skill-set. 



Research on postcolonial computing shows that mainstream computing knowledge is often ignorant towards local understanding of technologies, which creates a space of marginalization and failure against local communities \cite{ahmed2017digital, sultana2019witchcraft}. Localized knowledge and understanding of technologies have been historically marginalized, suppressed and neglected \cite{spivak1988can, sultana2019witchcraft}. This type of mindset represents a heightened danger for stigmatized individuals like Hijra. Even though Hijra are familiar with the basic utilization of the platforms, in many cases, they have no knowledge of the vast control settings that the platforms offer to them. For example, even though the usability of "disabling id" or “reporting” feature may seem straightforward to most users in Western context, due to the limited knowledge that is accessible to Hijra, participant X6 expressed frustration of not being able to identify and utilize the features during privacy concerns. Devito et. al's framework on social media ecosystem defines the online behavioral norms of LGBTQ+ communities from Western context by situating focus on their reliability on multiple spaces to share distinct content as a privacy control measure \cite{devito2018too}. Even though it touches based on how privacy controls on some individual platforms seemed to be irrelevant in the face of contextually- dependent behavior for their users, it did not factor the skill and knowledge that potentially dictate the relevancy of the privacy controls of a platform. Users can become unable to personalize or control the data they share online \cite{park2013digital}, as the task of exploring and mastering this protective technology is often left entirely to the adopters themselves \cite{atkinson2016breaking, ahmed2013ecologies}. 
For instance, for participants like P5, Facebook (with its huge investments into privacy tools) can be perceived as less safer and inclusive for presenting Hijra identity as opposed to other platforms due to P5's limited skill and knowledge about the features provided by the platform to report offensive anti-Hijra communities online. As Facebook relied on the assumption that users are skilled enough to adopt platform affordances, it potentially ignore marginalized populations like Hijra, who may not have the same access to the knowledge and skill. Being unable to utilize the privacy controls, many marginalized users face harassment online that effectively force them to restrict their participation on social media platforms \cite{turan2013reasons, nova2019online} or employ heavily tailored privacy settings to manage visibility within a single platform \cite{devito2018too}. For Hijra, they blame the space and redirect their participation elsewhere instead of utilizing the platform affordances. By accounting skill, we are able to directly interrogate the impact of the affordances on users' perceptions of online spaces, self-presentation and content sharing strategies. Therefore, this study suggests to include skill as a fourth element in the social media ecosystem framework that, beyond just Hijra or other GSM populations from certain contexts, potentially controls how users perceive online spaces and interact.

\subsubsection{Community and cultural influence on low skilled users}
While we establish technical skill or knowledge as a primary element within social media ecosystem for Hijra in non-Western context, we also observe community and cultural influence on these low skilled users that potentially construct their online practices as well. Hijra, with limited knowledge on platforms features, are often dependent on the information flow passed to them from their closely knitted community. Previous literature on information seeking states that users with limited resources seek and make sense of any information they receive when they have high-level self-presentation goals, but may not know exactly how to achieve them 
\cite{gillespie2014relevance, rubel2016black, devito2018people}. Connections like friends and families step in such cases and inform the users of their social media platforms' affordances, including lax privacy settings \cite{devito2018people}.
 However, for Hijra, such exogenous information come from other Hijra members of the community or sometimes from authoritative figures like Guruma, 
who often also have limited access to information on affordances. Their influences as sources of information on platform affordances impact the way Hijra's ecosystem are built or perceived. For example, participants like X6 (who is a Guruma) and P1 (fellow Hijra) are potentially motivated to direct social media participation based on the (limited) knowledge on platforms' affordances they have, whilst influencing other Hijra with limited skill set within the community to construct similar perception. 

As many Hijra get engaged with unsolicited work to earn money through distinct social media ecosystem, to protect themselves from unwanted harassment and invasion of personal privacy, Hijra are at constant look for adopting new social media platforms that will bring the more clients and better privacy. This adoption of new platforms often happens earlier with Gurumas who wish to ensure their followers safety online before they engage in it. 
It is fascinating to see how Hijra are not only influenced by their peers but also by their Gurumas who not only stipulate their physical lifestyle, but also dictate how their online participation should look like (often for their own safety). Taking influence of authoritative figure and community influence on limited skill-set into consideration help us to explore Hijra's perception, or in a broader sense GSM perception, on building their social media ecosystem more accurately. 

Apart from community influence, cultural influence on skill-set has also been an important factor for Hijra in Bangladesh. While many social media platforms offer its users extensive options of privacy settings, they are not always culturally appropriate and signified and thus, become unnecessary and useless. For example, even though participant P5 was struggling through intense online harassment for being Hijra online, it never occurred to them to "report" against someone or pages/groups, as the feature is not culturally signified to this low-skilled population. While previous research states that social media can reinforce personalized content curating behaviors within users that range from hiding posts to blocking other users \cite{zhu2017shield}, there some Hijra primarily online rely on the feature "block" than any other censorship tool to restrict harassment or unwanted connections. However, such feature of blocking users became impractical by the existence of fake profiles online. As Hijra often add unknown people to their existing online profiles to build connections, experiences of being deceived by the fake profiles make it harder for Hijra to trust the space. Trust and risk have been theorized as the most influential factors affecting individual adopting or rejecting behavior toward social media platforms \cite{wang2016understanding}. Whereas Hijra acknowledge harassing contents created against them are done by homophobic users in Bangladesh, they also count the space responsible for hosting such negative contents. Such distrust on a specific space influence Hijra to leave or shift their participation to another platform that is less harassing towards them or needs less technological understanding to control. This indicates an inadequacy within the platform design system that fails to serve GSM populations from a culturally different context.

\subsection{Implications for Design Practices}
Our findings add to the broader conversation on building more inclusive technologies for GSM populations by identifying specific design practices that need to be adopted by the designers during platform development. Be it for gender construction or privacy controls during online self-presentation, GSM users like Hijra from non-Western contexts struggle with platform designers' current design practices that are mostly informed by the Western notions of affordances \cite{sambasivan2018privacy, nemer2016online, ahmed2017digital}. Thus, to ensure more comprehensive online platforms for participants like Hijra, this study argues for incorporating non-Western marginalized views in design practices by the platform designers. This aligns with the recent body of work that signify the importance of developing appropriate technologies for non-marginalized as well as marginalized populations in both global north and south contexts \cite{nemer2013materializing, ahmed2017digital, devito2018too, haimson2015disclosure, bartsch2016control}. Our study highlights two of the most crucial implications for design practices in the context of non-Western GSM users that may help designers to develop more inclusive social media platforms. 


\subsubsection{Complicated Gender Constructions} First, due to the unfamiliarity with custom affordances that further requires users to fill out their own gender on social media, participants like X13, P43 in our study mentioned being unable to utilize platform's current mechanism to select gender that potentially forces them to compromise their identity construction. As the current design of most of the online platforms incorporate less granular gender choices for their users that only include male, female and custom/others options, this extra step to customize gender while not having hijra as a direct option similar to male/female categorization confuse Hijra, making their identity constructions on social media further challenging. Regarding this classification of gender provided by online platforms like Facebook, previous study mentioned that such classifications exists somewhere in-between a rigid binary and fluid spectrum \cite{bivens2017gender}. Even after Facebook has incorporated new ways to increase gender flexibility for its users, in their structural level, it still has continued to fit non-binary genders into binary classifications to serve stakeholders while also shaping the perceived needs and preferences of both users and advertisement clients \cite{bivens2017gender, bivens2016baking, scheuerman2019computers}. Similarly, Haimson and Hoffman have also questioned Facebook's underlying mechanisms used to impose legitimacy on its website, disproportionately affecting trans individuals and drag performers \cite{scheuerman2019computers, haimson2016constructing}. These prior studies back our arguments on the existing shortcomings of gender classification design practices of platforms like Facebook that need to be reevaluated and restructured from a non-Western perspective for communities such as Hijra. While Facebook prides to support GSM populations by providing 56 gender options (under custom) in the interface that pop up when a user attempts to type in their preferred gender term \cite{facebook_diversity_2014, bivens2017gender}, our study shows that for Hijra, it fails to support in similar manner. A user needs to type in the whole term "hijra"/"Hijra"/"Meye Hijra"/"Chele Hijra" to select their gender, whereas for other GSM users, such as trans populations, typing in only a letter brings up possible suggestions related to that gender \textbf{(see Fig. 9)}. 
\begin{figure}
\centering
  \includegraphics[width=.6\columnwidth]{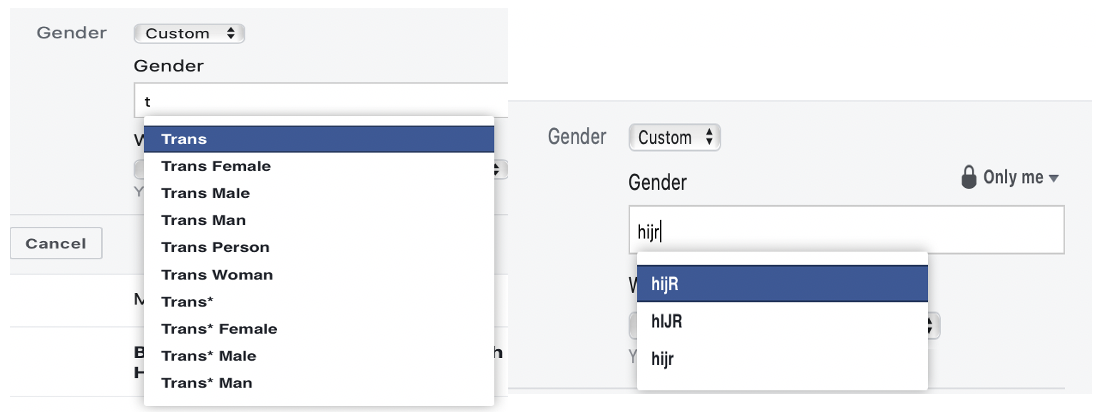}
  \caption{Classifications of Gender in Facebook}~\label{fig:figure11}
  \vspace{-2em}
\end{figure}This confirms how platform designers' understanding of gender is limited within a pre-set Westernized gender classifications and force GSM users from non-Western context to follow extra complicated steps to construct their identity online. This limitation within design practices brings forth a classic tension in information studies around classification and categorization where classifications provide orders but may miss out certain categories \cite{haimson2015user, bowker1999sorting}. While research like \cite{haimson2015disclosure} focused on similar constructions of gender through problematic design implications in Facebook, such research is predominantly based on Western trans populations who may have better understandings on platform affordances than Hijra from non-Western context. Thus, to set the focus more on non-Western GSM context, our study joins the discussion and encourages designers to think holistically about how these marginalized users prefer to define their gender online and provide more accessible and culturally appropriated options to them.

\subsubsection{Privacy Affordances} Second, our participants like P5 and P1 discuss Hijra's online struggle of using perceived platform affordances to combat community based online harassment that they face frequently due to their identity as Hijra. Even though platform designers include many privacy controls setting for their users while developing platforms, such as features like reporting or blockng individuals/disabling personal id, it was evident within our findings that such mechanism often do not work for Hijra from individual level. Our study finds, Hijra have strong community aspect within themselves; as such, their perceived platform affordances is significantly influenced by their group dynamics and largely dependent on the sharing of information within communities. Thus, incorporating and introducing group level privacy tools, such as collective block list, may benefit Hijra, as it can facilitate individual effort to combat online harassment with group support as well as can create a more manageable and user-friendly experience for them. Conceptualizations of online privacy remain mostly at the individual level in Western context \cite{belanger2011privacy, chen2013privacy, floridi2014open}, 
and while research like \cite{suh2018distinguishing, de2014managing, jhaver2018online} validates group privacy concerns as parts of design practices from a Western non-marginalized non-GSM context, our study strengthens these design implications by situating them in non-Western contexts for GSM. Of course, this must be weighed against technical understanding of the GSM users, as these design practices will end up adding more complexities to the existing privacy mechanism if the designers incorporate them within design practices without providing proper guidelines to their users who may lack access to the needed knowledge of using these tools. Therefore, platform designers need to think more expansively about how they can address this inaccessibility of knowledge and skill by the GSM users while also adding privacy flexibility to them from non-Western context, as it significantly impact their self-presentation online.

\section{Limitations and Future Work}
This study has some distinct limitations. Even though we wanted to ensure author accountability for the findings reported in this study by member checking our results with participants, we could not fulfill our objective. Due to our participants' extremely busy schedule and being severely affected by COVID-19 in Bangladesh, we could not meet our participants to discuss the reported results. Even though we accept it as an unfortunate limitation to our study design, we have tried our best to do extensive background research on Hijra community and carefully checked all the claims we made in the paper to ensure accountability. Additionally, while this study primarily focused on Hijra's self-presentation and online practices through social media ecosystem framework, it was outside the scope of this study to examine external circumstances such as local laws and changing political circumstances which can also effect Hijra's social status both offline and online. Future work could explore these structural factors while also investigating the possible intracommunity differences between trans and intersex Hijra individuals through an ecological lens \cite{devito2018too}, potentially yielding a more complete picture of Hijra, or in general GSM self-presentation in non-Western contexts.

\section{Conclusion}
Social interaction across multiple online platforms is a challenging issue for members of GSM due to the stigmatizations they face in daily basis for their identity, which increases the complexity of their self-presentation decisions. In this paper, through investigating personal social media ecosystem model, we have explored how GSM from non-western context, such as Hijra, construct their online participation and self-presentation focusing audiences, affordances and spaces in mind. This paper has also extended the model by incorporating user skill and knowledge as a fourth crucial element in the ecosystem that significantly impacts Hijra's perceptions on online space and online practices, such as audience management or platform migrations. Our contributions on community and cultural influence on Hijra's online participation have inspired us towards implications for design practices that
take account of more accessible and culturally appropriated gender categorizations for Hijra as well as group level privacy controls to facilitate these populations during online harassment. We believe these suggestions on design practices can provide guidance to the researchers and designers in further efforts to understand and support GSM in achieving their self-presentation goals, and lay the groundwork for future in-depth work on these populations in an increasingly diversifying social media space.

\section{Acknowledgement}
We would like to acknowledge \textit{Sachetan Somajsheba Hijra Sangha} (a non-profit organization) and its associated people for their enormous support in helping us being connected with Hijra communities for this project. Without their support, it would have been hard to conduct this study. We would also like to thank the guruma (requested to be unnamed) who has helped us during the data collection process and provided advice throughout the step. Furthermore, we would like to extend our gratitude to all of our reviewers who have provided us detailed and important suggestions during the review process that has helped us to build a much stronger submission. 

\bibliographystyle{ACM-Reference-Format}
\bibliography{sample-base}
\received{June 2020}
\received[revised]{October 2020}
\received[accepted]{December 2020}

\end{document}